\documentclass[12pt,letterpaper]{JHEP3}

\usepackage{slashed}
\usepackage{amsmath}
\usepackage{amssymb}
\usepackage{axodraw}

\newcommand{\ba}{\begin{eqnarray}}
\newcommand{\ea}{\end{eqnarray}}
\newcommand{\rmi}[1]{{\mbox{\scriptsize #1}}}

\newcommand{\tr}{{\rm tr\,}}
\newcommand{\nn}{\nonumber \\}
\newcommand{\fr}[2]{{\frac{#1}{#2}\,}}

\renewcommand{\(}{\left(}
\renewcommand{\)}{\right)}

\renewcommand{\ln}{{\rm ln}}

\def\half{{\textstyle \frac 12}}
\def\coeff#1#2{{\textstyle \frac {#1}{#2}}}
\def\Tc{T_{\rm c}}
\def\mD{m_{\rm D}}
\def\mE{m_{\rm E}}
\def\gE{g_3}

\def\x{\mathbf {x}}
\def\Z{\mathcal{Z}}
\def\zp{\bar z}
\def\vv{\bar v}
\def\openone{\rlap 1\kern 0.22ex 1}

\def\gauge{s}

\preprint{}
\title
    {
    \boldmath
    $Z(3)$-symmetric effective theory for
    $SU(3)$ Yang-Mills theory at high temperature
    }
\author
    {%
    A.~Vuorinen\footnote{\tt vuorinen@phys.washington.edu}\;
    and Laurence~G. Yaffe\footnote{\tt yaffe@phys.washington.edu}
    \\Department of Physics, University of Washington, Seattle, WA 98195--1560
    }%

\abstract
{
A three-dimensional effective theory for high temperature
$SU(3)$ gauge theory, which maintains the $Z(3)$
center symmetry of the full theory, is constructed.
Such a $Z(3)$ invariant effective theory
should be applicable to a wider temperature range than the
usual effective theory, known as EQCD,
which fails to respect the center symmetry.
This center-symmetric effective theory can reproduce
domain wall and phase transition properties
that are not accessible in EQCD.
After identifying a convenient class of $Z(3)$ invariant
effective theories,
we constrain the coefficients of the various terms
in the Lagrangian
using leading-order matching to EQCD
at high temperature,
plus matching of domain wall properties in the full theory.
We sketch the expected structure of the phase diagram of the effective theory
and briefly discuss the prospects of numerical simulations and the addition
of quarks.
}


\begin{document}


\section{Introduction}

The thermodynamics of high-temperature QCD with various numbers of
massless quarks has been widely studied using the method of dimensional
reduction, in which one exploits the decoupling of the non-static
degrees of freedom to build an effective three-dimensional theory
known as electrostatic QCD (EQCD) \cite{bn1}.
This approach has been used to derive the
perturbative expansion of the QCD pressure through order $g^6\,\ln\,g$,
and to compute various gluonic and mesonic correlation lengths
\cite{klry,hlp2,lv}.
It has also led to improved understanding of
the non-perturbative dynamics on the ``ultra-soft'' $g^2T$ energy scale
\cite{klry3}.
This approach is applicable
at asymptotically high temperatures where the running coupling
$g^2(T)$ is small.
But closer to the confinement/deconfinement transition,
at temperatures of a few times $\Tc$, even non-perturbative simulations of EQCD
have failed to produce satisfactory results \cite{klrrt, klry2}.
There have been attempts to build effective
theories for the Wilson line near the phase transition
\cite{pisa, kovner, peter, wiese},
but none of these theories (which often come with very complicated Lagrangians)
provide any connection to the perturbative regime at $T\gg \Tc$.

Our goal is to formulate a simple three-dimensional effective theory
for hot QCD that can reproduce equilibrium properties at temperatures
ranging from asymptotically large down to $\Tc$.
For simplicity,
most of our discussion will focus on the case of pure Yang-Mills theory,
although the addition of dynamical quarks
will be briefly discussed in the last Section.
The construction of our theory
relies on a separation between the inverse correlation length
and the lowest non-zero Matsubara frequency, $2\pi T$.
In $SU(3)$ Yang-Mills theory,
there is a parametrically large separation
between these scales at very high temperatures,
and a modest separation near $\Tc$.%
\footnote
    {
    It is natural to distinguish the correlation lengths
    of time-reflection even and odd operators.
    The Debye screening length $\mD^{-1}$ is the longest correlation
    length of time-reflection odd operators \cite{ay},
    such as the imaginary part of the trace of the Wilson line.
    The longest correlation length of time-reflection even operators,
    which we will denote by $\mu^{-1}$,
    is longer than the Debye length.
    In $SU(3)$ Yang-Mills theory at asymptotically high temperatures,
    $\mD/(2\pi T)$ approaches $g(T)/(2\pi)$
    while $\mu/(2\pi T)$ is approximately as $0.1 \, g(T)^2$
    \cite{owe}.
    (This is an estimate based on the corresponding $SU(2)$ result,
    plus the fact that $\mu$ should scale nearly linearly with $N_c$.)
    Correlators of purely magnetic operators fall off like
    $e^{-\mu |\x|}$ at large distance, while correlators of the
    real part of the trace of the Wilson line initially decrease like
    $e^{-2 \mD |\x|}$
    but eventually fall more slowly as $e^{-\mu |\x|}$.
    The correlator of the imaginary part of the trace of the Wilson line
    initially decreases like $e^{-3 \mD |\x|}$
    but switches to $e^{-\mD |\x|}$ at large distance.

    As the temperature decreases, the ratio $\mu/(2\pi T)$,
    as deduced from the Wilson line correlator,
    rises to about 0.5 at $T \sim 2 \Tc$
    \cite{lp,kacz}
    and then falls to about 0.1 just above $T_c$ \cite{kacz}.
    The ratio $\mD/(2\pi T)$ on the other hand rises to near unity at
    $T \sim 2 \Tc$ \cite{lp}.
    Because the confinement/deconfinement phase transition in $SU(3)$
    Yang-Mills theory is only weakly first order \cite{su3order},
    this ratio should decrease,
    similar to the behavior of $\mu/(2\pi T)$,
    closer to the phase transition,
    but we are unaware of any numerical simulations confirming this.
    Near $T_c$, correlations of purely magnetic operators
    appear to fall exponentially with a length scale near $(6T)^{-1}$
    but should eventually switch to the longer scale $\mu^{-1}$
    seen in the Wilson line correlator \cite{lp}.
    }

In pure $SU(3)$ Yang-Mills theory,
the confinement/deconfinement phase transition is a symmetry-breaking
phase transition.
The theory has a global $Z(3)$ symmetry corresponding to
invariance of the action under gauge transformations which are periodic in
the temporal direction only up to a twist belonging to the center of the
gauge group.
This $Z(3)$ symmetry is spontaneously broken in the deconfined phase where
the Wilson line operator (which is an order parameter for this symmetry)
acquires a non-zero expectation value.
Close to the phase transition region, fluctuations in the Wilson line
become increasingly important and the $Z(3)$ invariance of the Wilson line
probability distribution is an essential feature of the theory.
Even in full QCD with dynamical quarks it has been argued
that the $Z(3)$ symmetry,
although explicitly broken by the presence of light quarks,
may nevertheless play an important role in the dynamics near $\Tc$ \cite{pisa}.

The standard high temperature effective theory, EQCD,
cannot provide a smooth interpolation between asymptopia and $\Tc$.
This is evident even without considering numerical evidence,
as the construction of EQCD violates one of the central tenets
of effective field theory --- it does not respect
all the symmetries of the underlying theory.
EQCD breaks the $Z(3)$ center symmetry explicitly,
even in the absence of quarks.
Its leading order Lagrangian is obtained by expanding the one-loop
effective potential of the Wilson line \cite{gpy} around one of its
three degenerate minima.
Effects of the explicit $Z(3)$ breaking
are clearly visible in the phase diagram of the effective theory \cite{klrrt}.%
\footnote
    {
    Even though it has been argued that a ``partial dynamical restoration''
    of the $Z(3)$ symmetry takes place in EQCD \cite{klrrt},
    this theory does not have three distinct, degenerate,
    and physically equivalent equilibrium states
    which would signal a spontaneously broken $Z(3)$ symmetry.
    These aspects will be discussed in more detail in Sec.~5.
    }

An effective theory that preserves the $Z(3)$ symmetry can capture
the dynamics of thermal fluctuations (above $\Tc$) which create bubbles
sampling all three degenerate equilibrium states,
and thus should have a phase structure similar to that of the full
four-dimensional theory.
In particular, its phase diagram should contain a quadruple point, where all
three deconfined phases coexist with the confining phase,
thereby extending the description of the theory into a temperature range
inaccessible to EQCD.

Our guide during the process of constructing a better effective theory
for $N_\rmi{f}=0$ QCD
is the requirement that the new theory provide a smooth link between
asymptotically high temperatures, where perturbation theory is valid,
and $\Tc$, where a $Z(3)$ symmetry restoring phase transition occurs.
At high temperatures, its predictions must reproduce those of EQCD
(within its domain of applicability).
But a $Z(3)$ invariant effective theory should also reproduce
the thermodynamics of domain walls which interpolate between
different equilibrium states related by the $Z(3)$ symmetry
\cite{kap}.

We begin in Section~2 with a brief review of the relevant aspects of
the thermodynamics of $SU(3)$ Yang-Mills theory, including dimensional
reduction and the role of the Wilson line as an order parameter for
the confinement/deconfinement phase transition.
Section~3 presents the
general form of our effective theory.
We perform lowest order matching
to ordinary $N_\rmi{f}=0$ EQCD at high temperatures
in Section~4,
and also compute the semiclassical domain
wall profile in our effective theory.
The expected phase diagram of our theory
is discussed in Section~5, and compared to that of EQCD.
Possible future directions, including the addition of quarks,
are briefly discussed in the concluding Section~6.
Some details of the analysis of Section~3
are relegated to the appendices.

\section{Thermodynamics of \boldmath ${SU(3)}$ Yang-Mills}
\subsection{$Z(3)$ center symmetry}

The functional integral representation for the partition function
of $SU(3)$ Yang-Mills theory
involves an integral over all gauge fields which are periodic
in (Euclidean) time with period $\beta$.
In addition to invariance under gauge transformations which are periodic
in time, the action and the space of field configurations are invariant
under gauge transformations $g$ which are only periodic in time up to
a ``twist'' belonging to the $Z(3)$ center of the gauge group,
\begin{subequations}
\begin{align}
    &A_{\mu}(x)
    \rightarrow
    \gauge(x) \, (A_{\mu}(x)+i\,\partial_{\mu})\, \gauge(x)^{\dagger} \,,
    &\gauge(x)\in SU(3) \,,
\\
    &\gauge(x+\beta \, \hat e_t) = z \, \gauge(x) \,,
    &z\in Z(3) \,.
\end{align}
\end{subequations}
Under such a transformation, the trace of the Wilson line
(or Polyakov loop)
\begin{eqnarray}
    \tr \, \Omega(\mathbf{x})
    &\equiv&
    \tr \biggl\{ {\cal P} \, \exp
    \bigg[i \int_0^{\beta}\!\!{\rm d\tau} \> A_0(\tau,\mathbf{x})\bigg]
    \biggr\}
    \,,
\label{omega}
\end{eqnarray}
transforms in the fundamental representation of $Z(3)$,
\begin{eqnarray}
    \tr\, \Omega(\mathbf{x})&\rightarrow & z \, \tr\, \Omega(\mathbf{x}) \,.
\end{eqnarray}
The magnitude of the expectation value of $\tr\Omega(\mathbf x)$
may be interpreted physically as the exponential of the change in
free energy due to the addition of an infinitely heavy
fundamental representation test quark at position $\mathbf x$,
\begin{eqnarray}
    |\langle \tr\Omega(\mathbf{x}) \rangle|
    &=&
    e^{-\beta \, \Delta F_q(\mathbf x)} \,.
\end{eqnarray}
This expectation value vanishes in the confining low temperature phase,
where the free energy cost to introduce a single test quark is infinite,
but the expectation value is non-zero in the deconfined high temperature phase,
where the free energy cost of a test quark is finite.
Hence, the $Z(3)$ center symmetry is unbroken in the confining phase,
but is spontaneously broken in the deconfined phase, and
the expectation value of $\tr\Omega(\mathbf{x})$ is an order parameter
for the $Z(3)$ center symmetry.%
\footnote
    {
    Unlike more typical symmetry breaking phase transitions,
    the low-temperature (confining) phase is the disordered phase,
    while the high-temperature (deconfined) phase is the ordered phase.
    }

Symmetry considerations \cite{ys}, as well as lattice simulations
\cite{su3order},
show that four-dimen\-sional SU(3) Yang-Mills theory belongs
to the same universality class as the three-dimensional three-state
Potts model, and undergoes a weak first order phase transition at
the critical temperature.
At the phase transition, the theory has a quadruple point with four co-existing
equilibrium phases
(the three spontaneously broken deconfined states,
related by $Z(3)$ transformations, plus the confining phase).
No correlation lengths diverge as $T \to \Tc$.

The transition from the deconfined plasma above $\Tc$
to the confining glueball phase below $\Tc$
may occur through two mechanisms: bubble nucleation or complete wetting.
In the former, bubbles of the confining phase are formed via thermal
fluctuations inside the (supercooled) deconfined matter just below $T=\Tc$,
and as the temperature is lowered these bubbles grow,
eventually filling all space.
Alternatively,
complete wetting takes place if one begins with domain walls
separating deconfined phases with differing $Z(3)$
``magnetization''
({\em i.e.}, differing phases for $\langle \tr \Omega \rangle$).
Such a domain wall may be viewed as a
thin film of the confining phase residing between the two deconfined phases.
As the temperature is lowered below $\Tc$,
this film widens and splits the ``deconfined-deconfined''
domain wall into two ``confined-deconfined'' domain walls which repel,
leading to the expansion of the confining phase in the
direction perpendicular to the original domain wall.

Domain walls separating $Z(3)$ rotated deconfined phases
are topologically stable, non-perturbative objects which
play an important role in the dynamics of quarkless QCD
at temperatures just above $\Tc$.
At sufficiently high temperatures,
their properties may be computed using semi-classical techniques
\cite{kap}.
In hot QCD, the effective potential for the Wilson line only arises
at one loop order \cite{gpy}.
As a result, the Debye mass $\mD$ (which is determined by the curvature
of this potential at its minimum) is of order $g(T) T$,
while the energy density at the top of the barrier separating $Z(3)$ minima
is $O(T^4)$
(not order $T^4/g(T)^2$, as one might have expected for a semiclassical object).
The width of a domain wall is determined by the Debye screening length.
So high temperature domain walls have an $O(T^4)$ energy
density over a distance of order $\mD^{-1}$, yielding a tension which
scales as $T^3/g(T)$.
This parametric dependence must be reproduced
by any effective theory that claims to capture the
$Z(3)$ symmetry breaking physics of hot Yang-Mills theory.

\subsection{Dimensional reduction and EQCD}

Due to the compact (Euclidean) temporal direction in a finite-temperature
field theory,
all four-dimensional fields may
be decomposed into Fourier sums running over Matsubara frequencies.
Each Fourier component acts as a three-dimensional field,
with a mass (for bosonic fields) given by an integer multiple of $2\pi T$.
At sufficiently high temperatures,
all non-static modes act like heavy fields and decouple from
dynamics on length scales large compared to $1/T$.
The remaining static modes are the relevant degrees of freedom
for physics on longer distance scales, yielding an effective
three dimensional theory.
To lowest order, the form of the effective
theory may be found by integrating out the heavy fields explicitly \cite{gpy}.
At higher orders,
it is more efficient to start with the
most general form of the Lagrangian and
adjust its parameters by demanding that the
effective theory reproduce a minimal set of physical quantities which
are computable in both the effective and underlying theories \cite{bn1,bn0}.

In the case of $SU(3)$ Yang-Mills theory at high temperature,
the degrees of freedom of the effective theory correspond
to the static components of the gluon field $A_{\mu}$.
Integrating out non-static fluctuations generates an order $gT$
mass for the temporal component $A_0$, due to Debye screening.
In addition, self-interaction terms are generated for the $A_0$ field.
The resulting Lagrangian of $N_\rmi{f}=0$ EQCD reads%
\footnote
    {
    This form of the effective theory is valid for $SU(2)$ and $SU(3)$.
    With four or more colors, there is an additional
    $ \(\tr A_0^2\)^2 $ term.
    But for both $SU(2)$ and $SU(3)$, the two fourth-order terms
    are not independent, as $ \(\tr A_0^2\)^2 = 2 \, \tr (A_0^4) $.
    \label{fn:double-trace}
    }
\begin{equation}
    { \mathcal L}_{\rmi{EQCD}}
    =  
    \gE^{-2}
    \left\{
    \half \,\tr F_{ij}^2
    + \tr\!\big[(D_iA_0)^2\big]
    + m_{\rm E}^2 \, \tr ( A_0^2 )
    + \lambda_{\rm E} \, \tr (A_0^4)
    \right\}
    + \delta{\mathcal L}_\rmi{E} \,,
\label{LEQCD}
\end{equation}
with $\gE\equiv g(T)\sqrt{T}$ the three-dimensional gauge coupling
and $D_i \equiv \partial_i -i [A_i,\,\cdot\,]$
an adjoint representation covariant derivative.
The final term $\delta{\mathcal L}_\rmi{E}$ stands for higher dimension
operators whose effects, on length scales large compared to $T^{-1}$,
are subleading and only affect the thermodynamics at order $g^7$.
The operators in $\delta\mathcal L_{\rm E}$ will be irrelevant for our purposes.

A straightforward way to obtain the form of the above Lagrangian,
plus the leading-order values of its parameters,
is to consider the one-loop effective potential
for the Wilson line
$
    \Omega(\mathbf{x})
    \equiv
    {\cal P} \,
    e^{i \int_0^{\beta}d\tau \> A_0(\tau,\mathbf{x})}
$
in the full theory.
The resulting potential \cite{gpy} has
three degenerate minima related by transformations belonging
to the $Z(3)$ center of $SU(3)$.
Choosing to work around the minimum where the Wilson line expectation value
is real and positive
(and subtracting from the potential an $A_0^3$
contribution originating from fluctuations in the static
gauge field, which are not to be integrated out in the present case),
one obtains Eq.~(\ref{LEQCD}) as
the resulting effective potential for $A_0$
together with the lowest-order values for the effective theory parameters,
\begin{equation}
    m_{\rm E}^2 = g(T)^2 \, T^2 \,,
    \qquad
    \lambda_{\rm E}
    = \coeff{3}{4} \, g(T)^2/\pi^2 \,.
\end{equation}
To lowest order, the mass parameter $m_{\rm E}$
is the same as the physical Debye mass $\mD$.

This derivation also reveals the major shortcoming of the
effective theory: when constructing
the EQCD Lagrangian, one has chosen to work in the neighborhood
of $A_0 \approx 0$, corresponding to a real, positive Wilson line,
and thus loses the $Z(3)$ invariance of the original theory.
The other $Z(3)$ minima correspond to eigenvalues of $A_0$ near
$\pm \coeff 23 \pi T$,
and are outside the domain of validity of the effective theory.
At asymptotically high temperatures this is of little consequence for
most physical quantities, as
the probability of thermal fluctuations crossing the barriers
separating different minima is exponentially small
[with an exponent scaling as $1/g(T)$].
But fluctuations between the different minima
become crucial for the dynamics closer to $\Tc$,
and hence a non-$Z(3)$-invariant effective theory cannot properly
describe physics near the confinement/deconfinement phase transition.

\section{\boldmath $Z(3)$ invariant effective theory}

\subsection {Degrees of freedom}

To build a three-dimensional $Z(3)$ invariant effective theory,
the minimal degrees of freedom are the spatial gauge field
$\mathbf A(\x)$ and the Wilson line $\Omega(\x)$.
But because $\Omega(\x)$ is a unitary matrix,
a theory with polynomial interactions for the Wilson line
will not be perturbatively renormalizable.
For an effective theory which is only intended to be valid
below a UV cutoff, this is not a fundamental problem.
However, it is a serious practical nuisance.
It complicates perturbative matching to the underlying fundamental theory,
as well as matching between lattice and continuum regulated versions of the
effective theory.

Therefore, we take as our goal the construction of an effective
theory whose short distance behavior is as benign as possible.
To do so, we will replace the unitary matrix $\Omega(\x)$
by an unconstrained $3 \times 3$ complex matrix $\Z(\x)$.
This should be viewed as analogous to the relation between
non-linear and linear sigma models
(or between an Ising model and a double-well $\phi^4$ field theory).
One may view the complex field $\Z(\x)$ as the result of
applying a block-spin renormalization group transformation
which averages $\Omega(\x)$ over a small spatial region,
producing a result which is no longer unitary.

An arbitrary complex matrix may always be decomposed into a product
of unitary and hermitian matrices, so we may write
$\Z(\x) = \Omega(\x) \, H(\x)$
with $H(\x) = H(\x)^\dagger$.
One may regard the matrix $H(\x)$ as containing unphysical degrees of freedom
(not present in the underlying Yang-Mills theory) which we
have chosen to ``integrate in'' in order to construct
an effective theory with good short distance behavior.
We will arrange for these extra degrees of freedom to be ``heavy''
--- to have masses large compared
to the inverse correlation length.

We require that the effective theory Lagrangian be invariant under
three dimen\-sional $SU(3)$ gauge transformations,
\begin{equation}
    \begin{array}{l}
    \Z(\x) \to \gauge(\x) \, \Z(\x) \, \gauge(\x)^\dagger \,,
    \\[5pt]
    \mathbf A(\x)
    \to
    \gauge(\x) \, (\mathbf A(\x) + i \mathbf\nabla) \, \gauge(\x)^\dagger \,,
    \end{array}
\end{equation}
for $\gauge(\x) \in SU(3)$,
as well as global $Z(3)$ phase rotations,
\begin{align}
    \Z(\x) &\to {\rm e}^{2\pi i n/3} \, \Z(\x) \,, \kern 2cm
\label {eq:Z3trans}
\end{align}
for integer $n$.

Instead of using the polar decomposition of $\Z(\x)$ mentioned above,
it will be more convenient for our purposes to decompose this
field into a traceless part and its trace (times a unit matrix).
Therefore, we define
\begin{eqnarray}
    L(\x) &\equiv& \tr \Z(\x)  \,,\qquad
\\
\noalign{\hbox{and}}
    M(\x) &\equiv& \Z(\x) -\coeff{1}{3} \, \tr \Z(\x)\,\openone \,,
\\
\noalign{\hbox{so that}}
    \Z(\x) &=& M(\x) + \coeff 13 \, L(\x) \, \openone \,.
\end{eqnarray}

\subsection {Effective Lagrangian}

The Lagrange density for the effective theory will have standard derivative
terms plus a potential for the complex scalar $\Z$,
\begin{eqnarray}
    \mathcal{L}
    &=&
    \gE^{-2}
    \left\{\strut
    \half \, \tr  F_{ij}^2
    + \tr\! \left(D_i \Z^{\dagger}D_i\Z\right)
    + V(\Z)
    \right\},
\label{lageff2}
\end{eqnarray}
with $D_i \equiv \partial_i -i [A_i,\,\cdot\,]$
and $F_{ij} \equiv \partial_i A_j - \partial_j A_i - [A_i,A_j]$.%
\footnote
    {
    The freedom to make multiplicative rescalings of $\Z$
    is used to fix the coefficient of
    $\tr (D_i \Z^{\dagger}D_i \Z)$ in the Lagrangian (\ref {lageff2}).
    One could also add a $|\mathbf\nabla (\tr \Z)|^2$ term, as
    this preserves all the required symmetries.
    But this additional term may be eliminated with a
    $\Z \to \Z + \alpha (\tr \Z) \openone$ field redefinition.
    }
The $Z(3)$ invariant potential $V(\Z)$ will consist of two pieces,
\begin{equation}
    V(\Z) = V_0(\Z) + \gE^2 \, V_1(\Z) \,,
\label{eq:V}
\end{equation}
with
\begin{align}
    V_0(\Z)
    &=  
    c_1 \,\tr\!\big[\Z^{\dagger}\Z\big]
    + c_2 \({\rm det}\!\big[\Z\big] + {\rm det}\big[\Z^{\dagger}\big]\)
    + c_3\, \tr\!\left[ (\Z^{\dagger}\Z )^{2}\right] ,
\label{v0}
\\\noalign{\hbox{and}}
    V_1(\Z)
    &=  
    \tilde{c}_1 \, \tr\!\big[M^{\dagger}M\big]
    + \tilde{c}_2 \(\tr [M^3] + \tr [(M^{\dagger})^{3}]\)
    + \tilde{c}_3 \, \tr\!\big[(M^{\dagger}M)^2\big] \,.
\label{v1}
\end{align}
We will refer to $V_0(\Z)$ as the ``hard potential''
and $V_1(\Z)$ as the ``soft potential''.
This terminology is appropriate at high temperatures where $\gE^2/T$ is small,
in which case the effective masses arising from $V_0(\Z)$ will be large
compared to those arising from $\gE^2 \, V_1(\Z)$.

The hard potential $V_0(\Z)$
is composed of the three lowest-dimension operators
which (in addition to being $Z(3)$ invariant)
are invariant under the $SU(3)\times SU(3)$ transformation
\begin{equation}
    \Z(\x) \to A \, \Z(\x) B \,,
\label{eq:su3su3}
\end{equation}
for arbitrary $A,B \in SU(3)$.
Note that the free $\tr (\partial_i \Z^\dagger \partial_i \Z)$ kinetic term
is also invariant under this transformation.
A short exercise, presented in Appendix \ref{sec:potmin},
shows that $V_0(\Z)$ has extrema when $\Z = \lambda \, \Omega$
for $\Omega$ any element of $SU(3)$ and $\lambda$ a real root of
$2 c_3 \, \lambda^3 + c_2 \, \lambda^2 + c_1 \, \lambda = 0$.
The coefficient $c_3$ must be positive for stability.
We will require that $c_2 < 0$ and $c_2^2 > 9\, c_1 c_3$.
These conditions ensure that the global minimum corresponds
to the largest positive real root
which we will write, for later convenience,
as $\lambda = \coeff 13 \, v$ with
\begin{equation}
    v
    \equiv
    \coeff 34
    \biggl(
    \frac{-c_2 + \sqrt{c_2^2 - 8 c_1 c_3} }{c_3} \>
    \biggr) .
\label{eq:v}
\end{equation}
Therefore, $V_0(\Z)$ is minimized when $\Z$ equals
an arbitrary $SU(3)$ matrix times $v/3$.%
\footnote
    {
    Note that sending the coefficients $c_i$ to infinity,
    while holding their ratios fixed,
    would cause $V_0(\Z)$ to act like a delta-function constraint
    forcing the matrix $\Z$ to equal an $SU(3)$ element (up to a scale factor).
    The resulting effective theory would be
    a gauged non-linear sigma model for the Wilson line $\Omega(\x)$.
    But this limit would return us to a non-renormalizable
    effective theory, which we wish to avoid.
    }

The coupling constants in the soft potential $V_1(\Z)$
will be required to satisfy the inequalities
$\tilde c_3 > 0$ and
$\tilde{c}_1\tilde{c}_3>\tilde{c}_2^2$.
As shown in Appendix \ref{sec:potmin},
these conditions ensure that $V_1(\Z)$ is strictly
positive for non-zero $M$.
Hence, the soft potential $V_1(\Z)$ is minimized when $M = 0$,
or equivalently when $\Z$ is proportional to the identity matrix.
The overall factor of $\gE^2$ multiplying the soft potential
is designed to mimic the fact, noted earlier, that the
effective potential for the Wilson line in high temperature Yang-Mills
theory only arises at one loop order.

We have only included in the soft potential the three lowest dimension
$Z(3)$ invariant terms which can be built just from the traceless field $M$.
There are more $Z(3)$ invariant terms, with the same dimensions,
which we could have added --- for example, terms including factors
of the trace $L$.
However, if $\Z$ is restricted to the manifold of minima
of the hard potential $V_0$,
and so is proportional to an $SU(3)$ matrix, then there are only
three independent operators which are gauge invariant, $Z(3)$ invariant,
and at most fourth order in the field.
The three terms we have included in the potential (\ref{v1}) are
a convenient choice for these three independent terms.
As shown in the next section, these terms will be sufficient
for matching with the EQCD Lagrangian.%
\footnote
    {
    Had we chosen to work with a gauge group
    $SU(4)$ or higher,
    we would need to include a quartic
    double trace term in Eq.~(\ref{v1}) in order to reproduce the EQCD
    Lagrangian.
    This is unnecessary for $SU(3)$, due to the identity
    mentioned in footnote \ref {fn:double-trace}.
    }

The soft potential $V_1(\Z)$ lifts the continuous degeneracy at the minimum
of $V_0(\Z)$, leaving three discrete minima related by the $Z(3)$ symmetry.
If $\Z$ equals $\coeff v3$ times an element of the center of $SU(3)$,
then it will simultaneously minimize both the hard and soft potentials,
and hence will trivially minimize the combined potential
$V(\Z) = V_0(\Z) + \gE^2 \, V_1(\Z)$.

The theory defined by Eqs.~(\ref{lageff2})--(\ref{v1}) is a superrenormalizable
three-dimensional field theory.
The only renormalizations required
(in addition to vacuum energy subtraction)
are one and two loop adjustments
of the coefficients $c_1$ and $\tilde c_1$ of the quadratic terms.

To carry out perturbative matching, as discussed in the next section,
we will need the expansion of the Lagrangian (\ref{lageff2}) about the minima
of the potential.
In the vicinity of the minimum at $\coeff v3 \, e^{2\pi i n/3}$,
we may decompose $\Z$ as
\begin{equation}
    \Z = e^{2\pi i n/3}
    \left\{
    \coeff 13 v \, \openone
    + \gE \left[
    \coeff 1{\sqrt 6} \, (\phi + i\chi) \openone + (h + i a)
    \right]
    \right\} ,
\label {eq:decomp}
\end{equation}
where $\phi$ and $\chi$ are real numbers,
while $h$ and $a$ are traceless Hermitian matrices.
We will also rescale the three-dimensional gauge field
$\mathbf A \to \gE \, \mathbf A$ to switch to conventional
perturbative normalization of the gauge field.
Note that the fluctuation fields $\phi$, $\chi$,
$h$ and $a$, as well as the gauge field $\mathbf A$, will now have
canonical dimension $1/2$, as appropriate for a three dimensional theory.
Expressed in terms of these fluctuation fields,
the effective theory Lagrangian has the form
\begin{align}
    \mathcal{L}
    =  
    V_{\rm min}
    &+ \half\, \tr F_{ij}^2
    + \half \left[ \(\partial_i \phi\)^2 + m_\phi^2 \, \phi^2 \right]
    + \half \left[ \(\partial_i \chi\)^2 + m_\chi^2 \, \chi^2 \right]
\nonumber\\[5pt]
    &+ \tr\!\big[ (D_i \, h)^2 + m_h^2 \, h^2 \big]
    + \tr\!\big[ (D_i \, a)^2 \big]
    + V_{\rm int}(\phi,\chi,h,a) \,,
\label{lag}
\end{align}
where $D_i \equiv \partial_i -i \gE [A_i, \cdot \,]$
and $V_{\rm min}$
is the value of $V_0(\Z)/\gE^2$ at its minimum.

The interaction terms in $V_{\rm int}$ are suppressed by one or more
powers of $\gE$, and come from the soft potential plus the
parts of the hard potential which involve more than two fluctuation fields.
These terms are displayed explicitly in Appendix \ref{sec:potshift}.
The leading order masses induced by the hard potential are
\begin{equation}
    m_{\phi}^2 =  
    -2 c_1 - \coeff 13 v \, c_2 \,,\qquad
\label{mphi}
    m_{\chi}^2 =  
    -v  \, c_2 \,, \qquad
    m_{h}^2 =  
    -2 c_1 - \, \coeff 43 v \, c_2 \,,
\end{equation}
with $v$ given in Eq.~(\ref{eq:v}).
The field $a$ stays massless at this lowest order.
Note that $m_h^2 = m_{\chi}^2+m_{\phi}^2$,
so this mass is not independent of the other two.
The relations between the couplings in $V_0(\Z)$ and
the tree level masses (and $v$) may be inverted to give
\begin{equation}
    c_1 = \coeff 16 ({m_{\chi}^2-3m_{\phi}^2}) \label{c1} \,,\qquad
    c_2 = -{m_{\chi}^2}/v  \,,\qquad
    c_3 = \coeff 3{4} {(m_{\chi}^2+3m_{\phi}^2)}/v^2 \,.
\end{equation}
The tree-level heavy masses $m_{\phi}$ and $m_{\chi}$,
plus the value of $v$,
should be regarded as arbitrary parameters for now,
subject to the condition that $m_\phi > \coeff 13 m_\chi$.
This is equivalent to the inequality
$c_2^2 > 9 c_1 c_3$
which ensures that the global minima of the hard potential $V_0(\Z)$
lie at $\coeff 13 v \, e^{2\pi in/3}$.
The value of the potential (divided by $\gE^2$) at these minima is
$
    V_{\rm min} \equiv -\fr{v^2}{108}(9m_{\phi}^2-m_{\chi}^2)/ \gE^2
$.

\section{Parameter matching at high temperature}

\subsection{Reduction to EQCD}

We now want to deduce the constraints on the coefficients $\tilde c_i$
in the soft potential $V_1(\Z)$ which must be satisfied if the effective
theory is to reproduce the long distance dynamics of $SU(3)$
Yang-Mills theory at asymptotically high temperature.
In this regime, fluctuations on the scale of the temperature $T$
[as well as on the Debye scale $g(T) T$] are weakly coupled,
so perturbative matching techniques may be used.
The most efficient way to carry this out is to start with our
$Z(3)$ invariant effective theory
and integrate out the heavy fluctuation fields $\phi$, $\chi$ and $h$.
This will produce a non-$Z(3)$ invariant effective theory
only involving the light field $a$ (and the three dimensional gauge field),
which may be directly compared to EQCD.

To lowest non-trivial order, the result of
integrating out the heavy fields is a ``light'' effective theory
with Lagrange density%
\footnote
    {
    We omitted the unit operator
    from the original effective theory Lagrangian
    (\ref {lageff2}), as well as from (\ref{lageff3}),
    for simplicity of presentation.
    The coefficient of the unit operator gives the short distance contribution
    to the free energy density (divided by $T$).
    To match this contribution to the free energy,
    the coefficient of the unit operator in our effective theory
    must differ from the corresponding coefficient in EQCD
    by
    $
    - V_{\rm min}
    + \fr{1}{12\pi}(m_\phi^3 + m_\chi^3 + 8 m_h^3)
    $,
    up to higher order corrections suppressed by powers
    of $\gE^2$ divided by a heavy mass.
    \label{fn:unit op}
    }
\begin{eqnarray}
    \mathcal{L}_{\rm light}
    &=&
    \half \, \tr  F_{ij}^2
    + \tr\! [(D_i \, a)^2 + m_a^2 \, a^2 + \tilde\lambda \, a^4] \,.
\label{lageff3}
\end{eqnarray}
Except for the trivial rescaling $\mathbf A \to \gE \, \mathbf A$
of the gauge field, and renaming $A_0 \to \gE \, a$,
this has the same form, and identical derivative terms,
as the EQCD Lagrangian (\ref {LEQCD}).

\begin{FIGURE}[ht]
{
\centerline{\epsfxsize=13.5cm\epsfysize=4.5cm\epsfbox{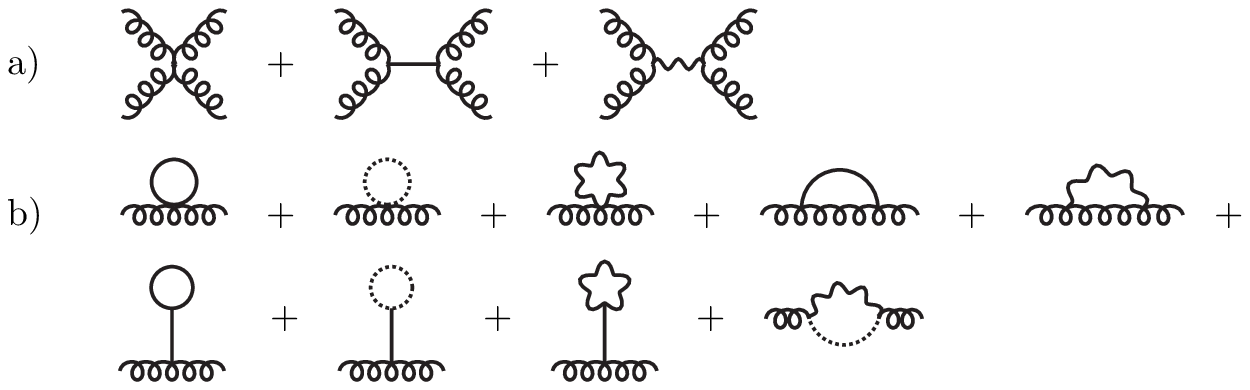}}
\caption[a]
    {
    (a)
    Tree-level diagrams
    originating from the hard potential $V_0$
    which contribute to the quartic interaction vertex
    for the light field $a$.
    (b) One-loop diagrams produced by $V_0$ which
    contribute to the mass of the field $a$.
    Solid, dotted, wavy and curly lines denote the propagators of
    the $\phi$, $\chi$, $h$, and $a$ fields, respectively.
    \label{fig:match}
    }
}
\end{FIGURE}

The effective light mass $m_a^2$,
and the quartic coefficient $\tilde\lambda$,
can have contributions arising solely from the hard potential $V_0(\Z)$,
as well as having contributions which involve the soft potential $V_1(\Z)$.
The lowest order contributions generated by the hard potential are
illustrated in Fig.~\ref{fig:match}.
Using the coefficients of the various terms in the expansion of
$V_0$ about its minima, as given in Eq.~(\ref{eq:Vshift}),
one may easily evaluate these diagrams explicitly.
However, this is not necessary --- the diagrams shown in
Fig.~\ref{fig:match}, at zero external momentum, sum to zero.
This is a consequence of the $SU(3)\times SU(3)$ invariance (\ref {eq:su3su3})
which the effective theory would have in the
absence of the soft potential $V_1(\Z)$ and the static gauge field.
An infinitesimal $SU(3)\times SU(3)$ transformation acts as a shift
on the light field $a$.
Integrating out the heavy fields,
without the soft potential,
must produce a result which respects
this invariance, and thus can only involve
derivative interactions for the light field $a$.

Consequently, the light mass $m_a^2$ and the quartic coupling $\tilde\lambda$
in the Lagrangian (\ref {lageff3})
can arise only from contributions involving the non-$SU(3)\times SU(3)$
invariant soft potential $V_1(\Z)$.
To lowest order, one may simply read off their values from the
soft terms in the shifted potential (\ref{eq:Vshift}), giving
\begin{equation}
    m_a^2 = \tilde c_1 \, \gE^2 \,, \qquad
    \tilde\lambda = \tilde c_3 \, \gE^4 \,.
\label {eq:lightcoeffs}
\end{equation}
Demanding that these coefficients coincide with the corresponding terms
in the EQCD Lagrangian (\ref {LEQCD})
(after identifying $A_0 \to \gE \, a$) fixes
\begin{eqnarray}
    \tilde{c}_1&=&T + {\mathcal O}(\gE^2) \,,\label{ct1}
\\
\noalign{\hbox{and}}
    \tilde{c}_3&=&\fr 3{4 \pi^2 T} +{\mathcal O}\Bigl(\frac{\gE^2}{T^2}\Bigr)
    \,,\label{ct3}
\end{eqnarray}
while $\tilde{c}_2$ is left undetermined.

Integrating out the heavy fields beyond leading order will, of course,
produce corrections to the coefficients (\ref {eq:lightcoeffs}),
suppressed by additional powers of $\gE^2$ divided by a heavy mass,
as well as induce additional higher dimension operators in the
light effective Lagrangian (\ref {lageff3}).
The first new operators, involving six powers of the light field $a$, will have a
coefficient of order $(\gE^2/m_{\rm heavy})^4$.

\subsection{Domain wall properties}

Having ensured that our theory reproduces the perturbative physics of
$SU(3)$ Yang-Mills theory at very high temperature,
we are left with two undetermined parameters, $v$ and $\tilde{c}_2$
(in addition to the deliberately introduced heavy masses $m_\phi$ and $m_\chi$).
As we will show, these two parameters may be fixed by
demanding that our effective theory reproduce the tension and
width of domain walls separating different $Z(3)$ minima.

Consider a domain wall lying in the $x$-$y$ plane,
which interpolates between the state with
$\langle \tr \Z \rangle = v$ at $z = -\infty$ and
$\langle \tr \Z \rangle = e^{2\pi i/3} \, v$ at $z = +\infty$.
To find the domain wall profile and evaluate its tension,
we need to minimize the free energy
for configurations of the scalar field
$\Z(z)$ which depend on a single coordinate $z$ and
satisfy the boundary conditions
\begin{subequations}
\begin{eqnarray}
    L(z=-\infty)&=&v \,,\qquad
    L(z=\infty)= e^{2\pi i/3} \, v \,, \\
    M(z=-\infty)&=&0 \,,\qquad
    M(z=\infty)=0 \,.
\end{eqnarray}
\end{subequations}
To evaluate the (leading order) domain wall tension
in the high temperature or weak coupling limit
it will be necessary to include:
\begin{enumerate}
\item
    The contribution of the hard potential,
    $\frac T{\gE^{2}} \int dz \> V_0(\Z(z))$.
\item
    The gradient energy,
    $\frac T{\gE^{2}} \int dz \> \tr |D_z \Z(z)|^2$.
\item
    The contribution of the soft potential,
    $T \int dz \> V_1(\Z(z))$.
\item
    The one-loop contribution from fluctuations in the gauge and scalar fields.
\end{enumerate}
Our task is greatly simplified by the hierarchical structure
of the potential.
Because of their overall factors of $\gE^{-2}$, one might expect the
first two contributions to be most important.
However, a translationally non-invariant domain wall configuration
can minimize $V_0(\Z)$, everywhere in space, provided
$\Z(z)$ equals $v/3$ times some (spatially varying) $SU(3)$ matrix.
And the gradient energy will be comparable to the energy due to
the soft potential, and not parametrically larger,
if the width of the domain wall is inversely proportional to $\gE$.
Because the soft potential is multiplied by an explicit factor of $\gE^2$
in the Lagrangian (\ref {lageff2}),
its contribution is comparable to the effects due to fluctuations in
the fields.
Hence, the one-loop contribution cannot be neglected.%
\footnote
    {
    This parallels the computation of the leading order
    domain wall tension in the full theory \cite {kap}.
    }

Using gauge invariance,
we may diagonalize $\Z(z)$ and write it as
\begin {equation}
    \Z(z) = {\rm diag} (f_1(z), f_2(z), f_3(z)) \,,
\end {equation}
with
\begin{subequations}
\begin{eqnarray}
    f_1(z)&\equiv& \coeff v3 \, e^{\fr{2\pi i}{3}[\alpha(z)+3\beta(z)]} \,,
\\
    f_2(z) &\equiv&  \coeff v3 \, e^{\fr{2\pi i}{3}[\alpha(z)-3\beta(z)]} \,,
\\
    f_3(z) &\equiv&  \coeff v3 \, e^{\fr{-4\pi i}{3} \alpha(z)} \,.
\end{eqnarray}
\end{subequations}
Here $\alpha$ and $\beta$ are real functions of $z$ that satisfy
the boundary conditions
\begin{subequations}
\label{eq:bc}
\begin{eqnarray}
    \alpha(z=-\infty)&=&0 \,,\qquad \alpha(z=\infty)=1, \\
    \beta(z=-\infty)&=&0 \,,\qquad \beta(z=\infty)=0 \,.
\end{eqnarray}
\end{subequations}
Note that shifting $(\alpha,\beta)$ by any integer multiple
of $(\coeff 32, \pm \coeff 12)$ leaves $\Z(z)$ invariant
(up to a permutation of its diagonal elements).

It will be convenient to introduce a dimensionless rescaled
coordinate $\zp$ and Wilson line expectation value $\vv$ via
\begin{eqnarray}
    \zp \equiv \gE \sqrt T \, z\, ,\qquad\vv \equiv \fr{v}{T}\, .
\end{eqnarray}
The gradient and soft potential contributions to the
domain wall free energy (per unit area) then become
\begin {equation}
    F_{\rm grad}
    +
    F_{\rm soft}
        =\gE^{-1} \, (\pi \vv \, T)^2 \, (\coeff 23\sqrt{T})^3 \,
    \int_{-\infty}^\infty
    d\zp \left\{ (\alpha')^2 + 3 (\beta')^2 + U_1(\alpha,\beta) \right\} ,
\end {equation}
with $\alpha' \equiv \partial \alpha / \partial \zp$, {\em etc.}, and
\begin {eqnarray}
    U_1(\alpha,\beta)
    &=&
    \fr{1}{4\pi^2}
    \Big\{
    3 - \cos(4\pi\beta)
    - 2 \cos (2\pi\alpha)\, \cos(2\pi\beta)
    \Big\}
\nn[5pt]
    &+&
    \fr{\vv \tilde c_2}{18 \pi^2}
    \Big\{
    6+\cos(4\pi\alpha)-3\cos(4\pi\beta)
    +2\cos(2\pi\alpha)
    \bigl[\cos(6\pi\beta) - 3\cos (2\pi\beta)\bigr]
    \Big\}
\nn[5pt]
    &+&
    \!\Bigl(\frac{\vv}{12\pi^2}\Bigr)^2
    \Big\{
    9+\cos(8\pi\beta)
    +2\cos (4\pi\beta) \bigl[ \cos(4\pi\alpha) {-} 2 \bigr]
    -8\cos (2\pi\alpha)\, \cos(2\pi\beta)
    \Big\} .
\nn
\end {eqnarray}

Fluctuations of the fields in the effective theory,
at one-loop order, produce functional determinants
depending on the background domain wall profile $\Z(z)$.
The width of the domain wall will be parametrically large compared to $1/v$,
and therefore one may use a gradient expansion in the evaluation of the
resulting functional determinants.
It is sufficient to keep just the first term, leading to%
\footnote
    {
    The evaluation of this one-loop contribution
    is similar to the calculation in Ref.~\cite{kap}.
    One may neglect the soft potential $V_1(\Z)$.
    Transverse fluctuations of $\mathbf A$
    acquire an effective mass due to the background value
    of the scalar field, and yield the above functional determinant.
    If one uses the gauge fixing term $\tr (\partial_i A_i)^2$
    and lets $\Z(x) = (\openone + \gE \, \eta(x)) \, \bar \Z$,
    with $\bar \Z$ the background value of the scalar field,
    then the longitudinal part of $\mathbf A$ mixes with the
    traceless anti-Hermitian part of $\eta$ (which receives
    no tree-level mass from the hard potential $V_0$).
    The resulting functional determinant is independent of $\bar\Z$,
    and equals one with dimensional regularization.
    Fluctuations of the trace and traceless Hermitian parts of $\eta$
    are also independent of $\bar\Z$, and produce an irrelevant
    constant (given in footnote \ref{fn:unit op}).
}
\begin {eqnarray}
    \frac{F_{\rm fluc}}{T}
    &=&
    \ln\det\!\left(
    -\partial^2 + \big[\Z^{\dagger},\big[\Z,\,\cdot\big]\big]
    \right) / \mathrm {(Area)}
\nonumber
\\
    &=&
    \sum_{ij}
    \int dz
    \int\fr{{\rm d}^3k}{(2\pi)^3} \> \ln\(k^2+|f_i(z)-f_j(z)|^2\)
    \nn &=&
    -
    \int \frac {dz}{3\pi} \>
    \( |f_1-f_2|^3+|f_2-f_3|^3+|f_3-f_1|^3 \)
    \nn &=&
    \gE^{-1} \, \pi^2 \vv^2 T \, (\coeff 23 \sqrt T)^3
    \int_{-\infty}^\infty
    d\zp \; U_2(\alpha,\beta) \,,
\label {eq:Ffluc}
\end{eqnarray}
with
\begin{equation}
    U_2(\alpha,\beta)
   = -\fr {\vv}{3\pi^3 }
    \Big\{
        \bigl|\sin (2\pi\beta)\bigr|^3
        + \bigl|\sin (\pi(\alpha{-}\beta))\bigr|^3
        + \bigl|\sin (\pi(\alpha{+}\beta))\bigr|^3
    \Big\} .
\end{equation}
Combining the three contributions gives
\begin{eqnarray}
    F_{\rm dw}[\alpha,\beta]
    &\equiv&
    F_{\rm grad} + F_{\rm soft} + F_{\rm fluc}
\nn &=&
    \gE^{-1} \, (\pi \vv \, T)^2 \, (\coeff 23 \sqrt T)^3
    \int_{-\infty}^\infty \! d\zp
    \left[
    (\alpha')^2 + 3(\beta')^2
    + U_1(\alpha,\beta)
    + U_2(\alpha,\beta)
    \right] \!.\qquad
    \label{lageff}
\end{eqnarray}
This is the excess free energy density due to a domain wall
(relative to the spatially constant equilibrium free energy density).
Note that all three contributions are parametrically the same size,
as anticipated.
Minimizing this functional, subject to the boundary conditions
(\ref {eq:bc}),
will yield the leading order domain wall tension.

\begin{FIGURE}[t]
{
\centerline{\epsfxsize=7.0cm\epsfysize=7.0cm\epsfbox{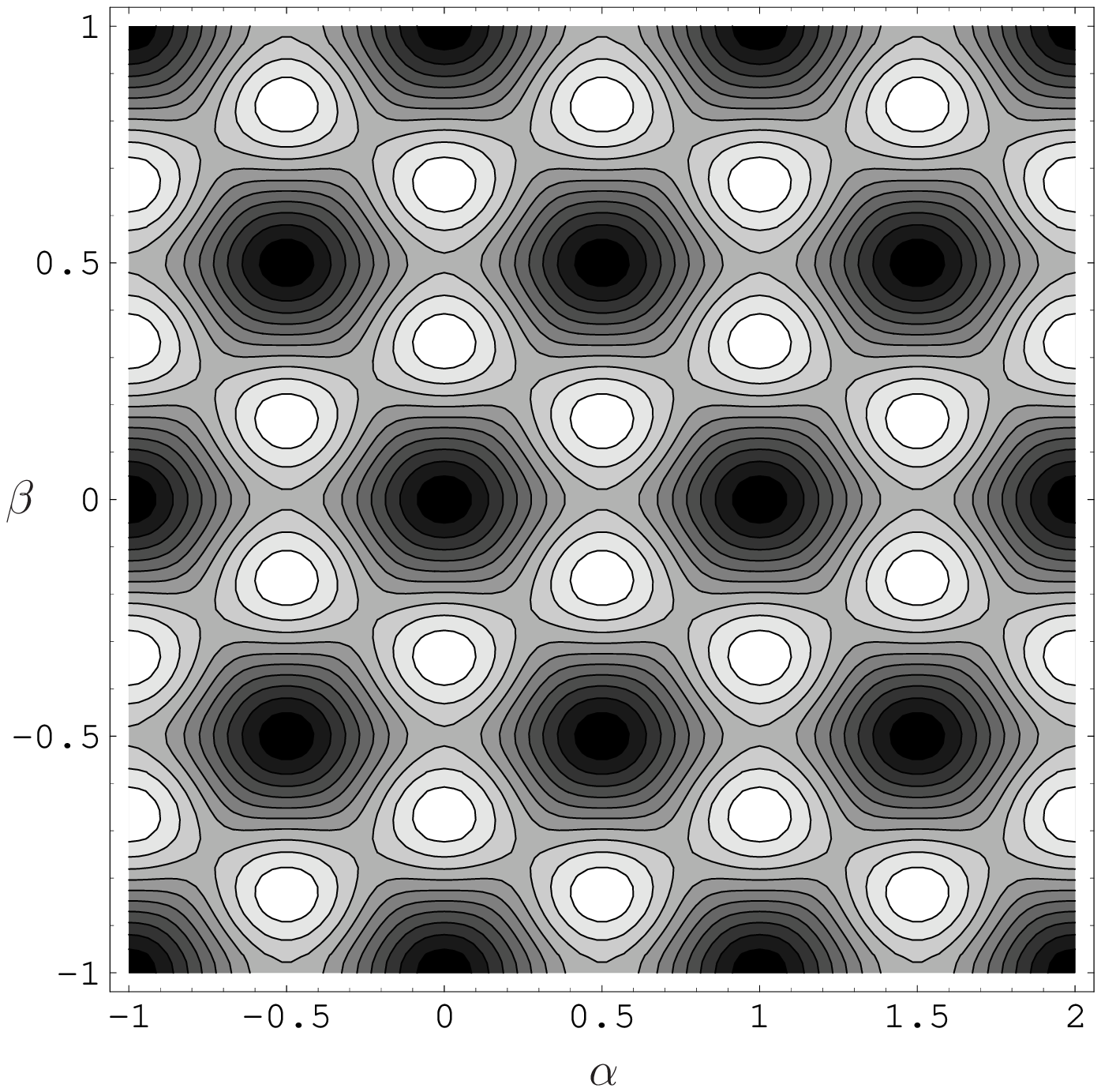}\epsfxsize=8.5cm\epsfysize=7.0cm\epsfbox{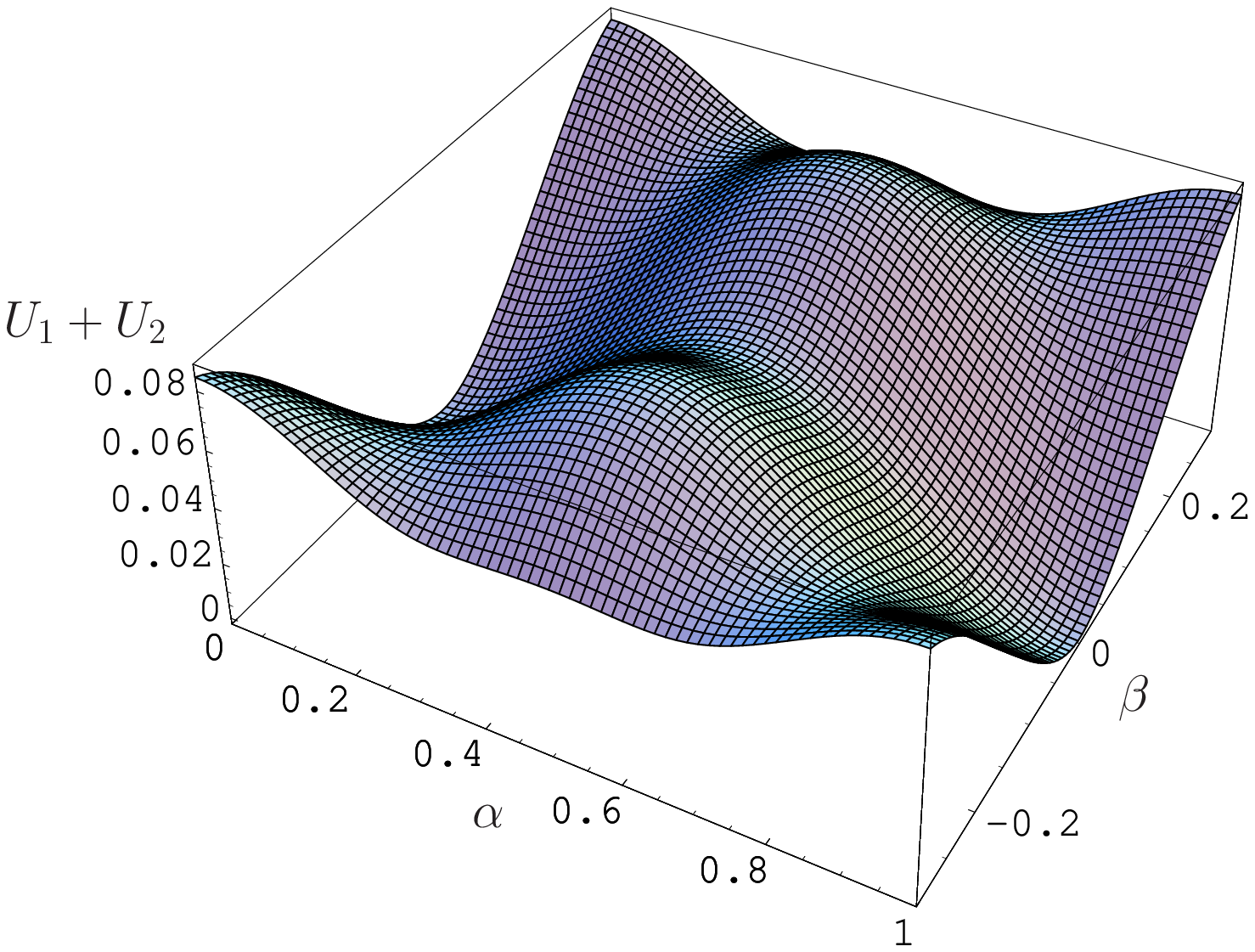}}
\caption[a]%
{The potential $U_1(\alpha,\beta) + U_2(\alpha,\beta)$ plotted for
the optimal values of $\vv$ and $\tilde c_2$ given in
Eq.~(\ref{ct2}).
In the contour plot on the left, minima correspond to the dark regions.
A fundamental domain,
representing all configurations of $\Z$ which are
inequivalent under conjugation by $SU(3)$,
is given by the triangle with vertices at
($\alpha,\beta$) = $(0,0)$, $(1,0)$ and $(1/2,1/2)$ --- which
represent the three distinct minima related by $Z(3)$ transformations.
The $3d$ plot on the right shows the region between the two minima at
$(\alpha,\beta) = (0,0)$ and $(1,0)$.
\label{fig:pot}}
}
\end{FIGURE}

The combined potential $U_1(\alpha,\beta) + U_2(\alpha,\beta)$
is plotted in Fig.~\ref{fig:pot}.
From the general form of the potential
it is easy to see that for positive $\tilde{c}_2$,
and all values of $\alpha$,
there is a local minimum in the $\beta$-direction at $\beta=0$.
This suggests, and further inspection confirms,
that one may simply set $\beta=0$ in the search for the minimum.
This reduces the computation to the minimization of
\begin{eqnarray}
    F_{\rm dw}[\alpha,0]
        =\gE^{-1} \, (\pi \vv\, T)^2 \, (\coeff 23 \sqrt T)^3
    \int_{-\infty}^\infty
    d\zp \> \big[ (\alpha')^2 + \widetilde U(\alpha) \big] ,
    \label {eq:Fdw}
\end{eqnarray}
with
\begin{eqnarray}
    \widetilde U(\alpha)
    \equiv
      \fr 1{\pi^2} \sin^2(\pi\alpha)
    - \fr {2 \vv}{3 \pi^3 } \sin^3(\pi \alpha)
    + \fr { {\vv}^2 +4\pi^2{\vv} \, \tilde{c}_{2} }{9\pi^4}
        \sin^4(\pi\alpha)
    \,.
\end{eqnarray}

Solutions to the equation of motion produced by varying
the functional (\ref {eq:Fdw}) conserve the ``energy''
$
    E \equiv \(\partial_{\zp} \alpha\)^2 - \widetilde{U}(\alpha)
$,
and the domain wall boundary conditions (\ref {eq:bc})
imply that $E = 0$.
The free energy functional (\ref {eq:Fdw}) is
invariant under the transformation
$\alpha(\zp) \to 1-\alpha(-\zp)$,
which is a combination of parity, charge conjugation,
and a $Z(3)$ symmetry transformation.
The minimal energy domain wall solution should have the same invariance,
implying that
$
    \alpha(\zp{=}0) = \half
$.
Hence, the appropriate solution to the zero energy
equation of motion
$
    \partial_{\zp} \alpha = \widetilde{U}(\alpha)^{1/2}
$
is given by
\begin{eqnarray}
    \int_{1/2}^{\alpha(\zp)} d\alpha \; \widetilde{U}(\alpha)^{-1/2}
        =\zp \, ,
\label{eom1}
\end{eqnarray}
while the resulting domain wall tension is
\begin{eqnarray}
    \sigma
    &\equiv&
    F_{\rm dw}[\alpha,0]
        =\gE^{-1} \, (4 \pi \vv \, T)^2 \, (\coeff 13 \sqrt T)^3
    \int_{0}^{1}\! d\alpha \> {\widetilde{U}(\alpha)^{1/2}} \,.
\label {eq:tension}
\end{eqnarray}
We will define
the width $\Delta z$ of the domain wall as
the ratio of the first moment of the domain wall free energy density
divided by the tension,%
\footnote
    {
    Alternative definitions of the domain wall width could, of course,
    be chosen, but make little difference.
    For example, defining the width from the second moment of the
    domain wall free energy density changes the resulting optimal
    parameters (\ref {ct2}) by about 0.1\% for $\vv$ and 2.6\% for $\tilde{c}_2$.
    }
or
\begin {eqnarray}
    \Delta z
    &\equiv&
    \frac {(4\pi \vv)^2 \, (\coeff 13 \, T)^3} {\gE^2 \, \sigma}
    \int_{-\infty}^\infty d\zp \> |\zp| \> \widetilde U(\alpha(\zp))
\nonumber\\[5pt]
    &=&
    \frac {2(4\pi \vv)^2 \, (\coeff 13 \, T)^3} {\gE^2 \, \sigma}
    \int_{1/2}^1 d\alpha_1 \> {\widetilde U(\alpha_1)^{1/2}}
    \int_{1/2}^{\alpha_1} d\alpha_2 \> {\widetilde U(\alpha_2)}^{-1/2}
    \,.
\label {eq:width}
\end {eqnarray}

\begin{FIGURE}[ht]
{ \centerline{\def\epsfsize#1#2{0.6#1}\epsfbox{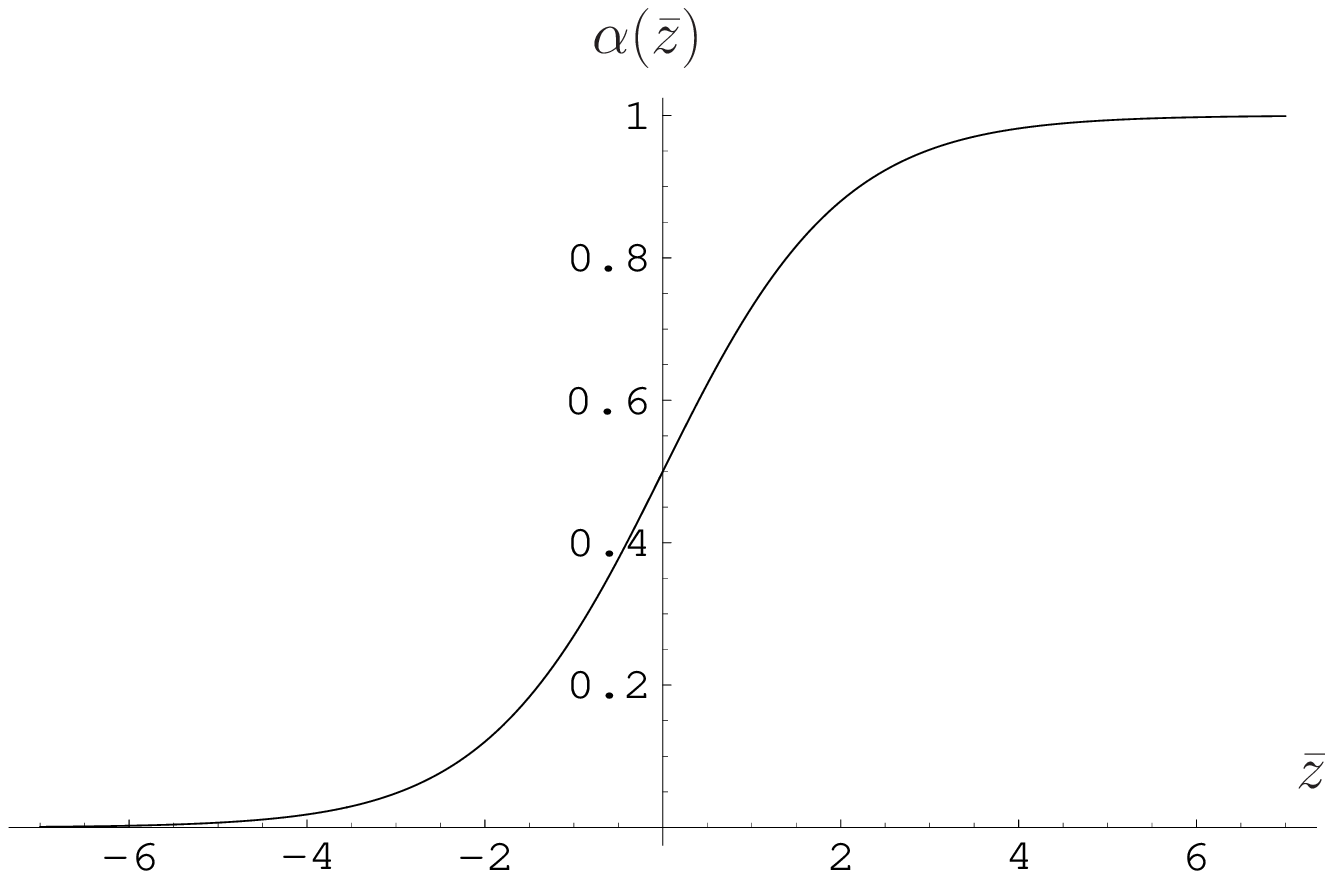}}
\caption[a] { The function $\alpha(\zp)$ plotted for the optimal
values of $\vv$ and $\tilde{c}_2$ given in Eq.~(\ref{ct2}).
 \label {fig:soln} } }
\end{FIGURE}%

None of the integrals (\ref {eom1})--(\ref {eq:width})
are analytically tractable
but evaluating them numerically,
or integrating the equation of motion
$
    \partial_{\zp} \alpha = \widetilde{U}(\alpha)^{1/2}
$,
for particular values
of $\tilde{c}_2$ and $\vv$ is straightforward.
The resulting tension and width may then be compared to
the corresponding results in the underlying Yang-Mills theory.
The (leading order) domain wall free energy density in
high temperature $SU(3)$ Yang-Mills theory is
\cite{kap}
\begin{eqnarray}
    \mathcal F_{\rm YM}(z)&=& \fr{\pi^2 \, T^4}{3\cosh^{4}(\zp(z)/2)} \,.
\label{eq:FYM}
\end{eqnarray}
The resulting domain wall tension is
\begin{eqnarray}
    \sigma_\rmi{YM}
    &=&
    \int_{-\infty}^\infty dz \> \mathcal F_{\rm YM}(z)
    = 
    \fr{8\pi^2}{9}\fr{T^3}{g(T)} \,,
\end{eqnarray}
and the (first moment) domain wall width equals
\begin{equation}
    \Delta z_{\rm YM}
    = 
    \sigma_{\rm YM}^{-1}
    \int_{-\infty}^\infty dz \> |z| \, \mathcal F_{\rm YM}(z)
    = 
    \fr{\ln (4) - \half}{g(T) \, T} .
\end{equation}

\clearpage

\begin{FIGURE}[ht]
{
\centerline{\def\epsfsize#1#2{0.55#1}
    \epsfbox{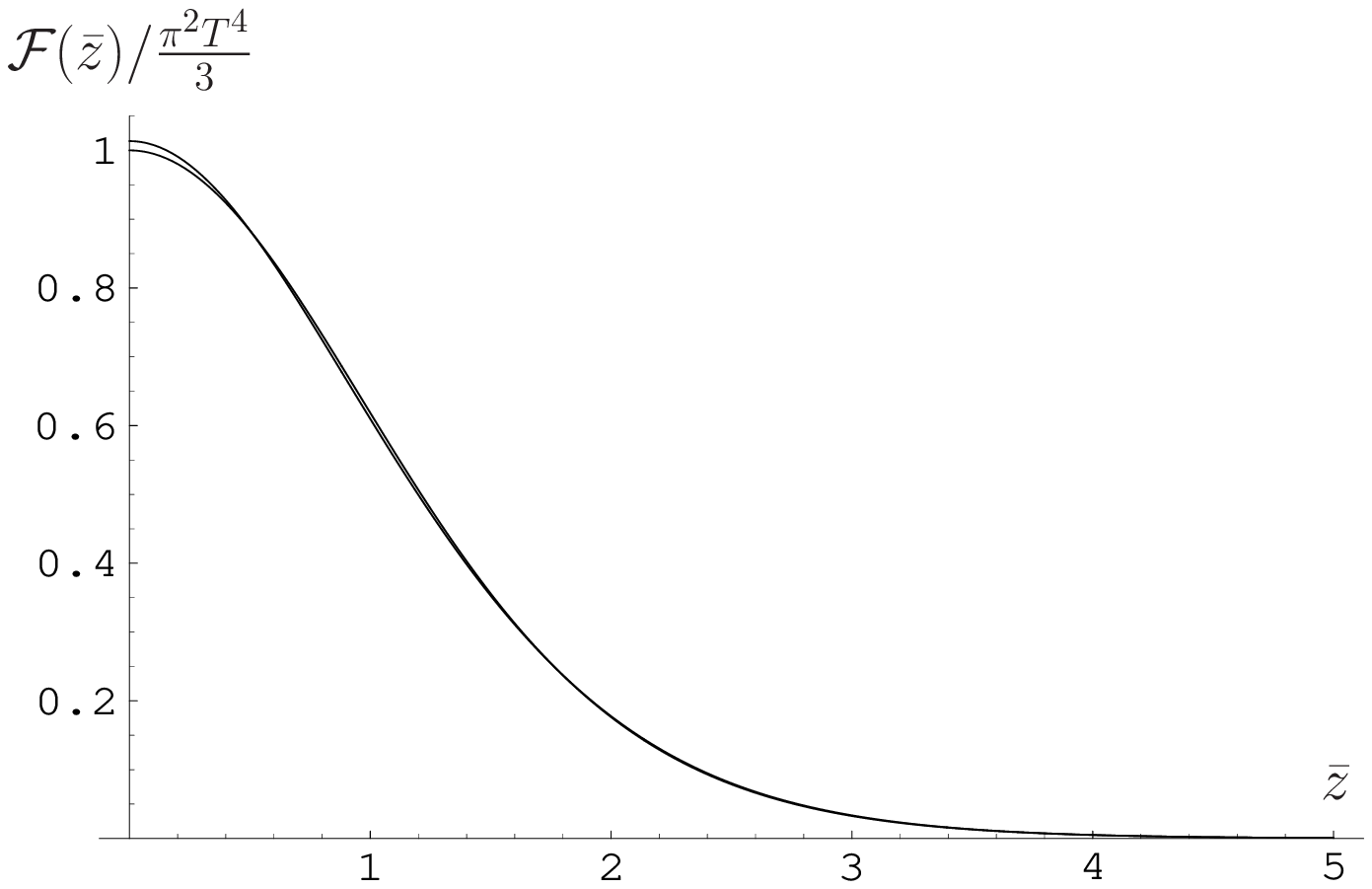}
    \epsfbox{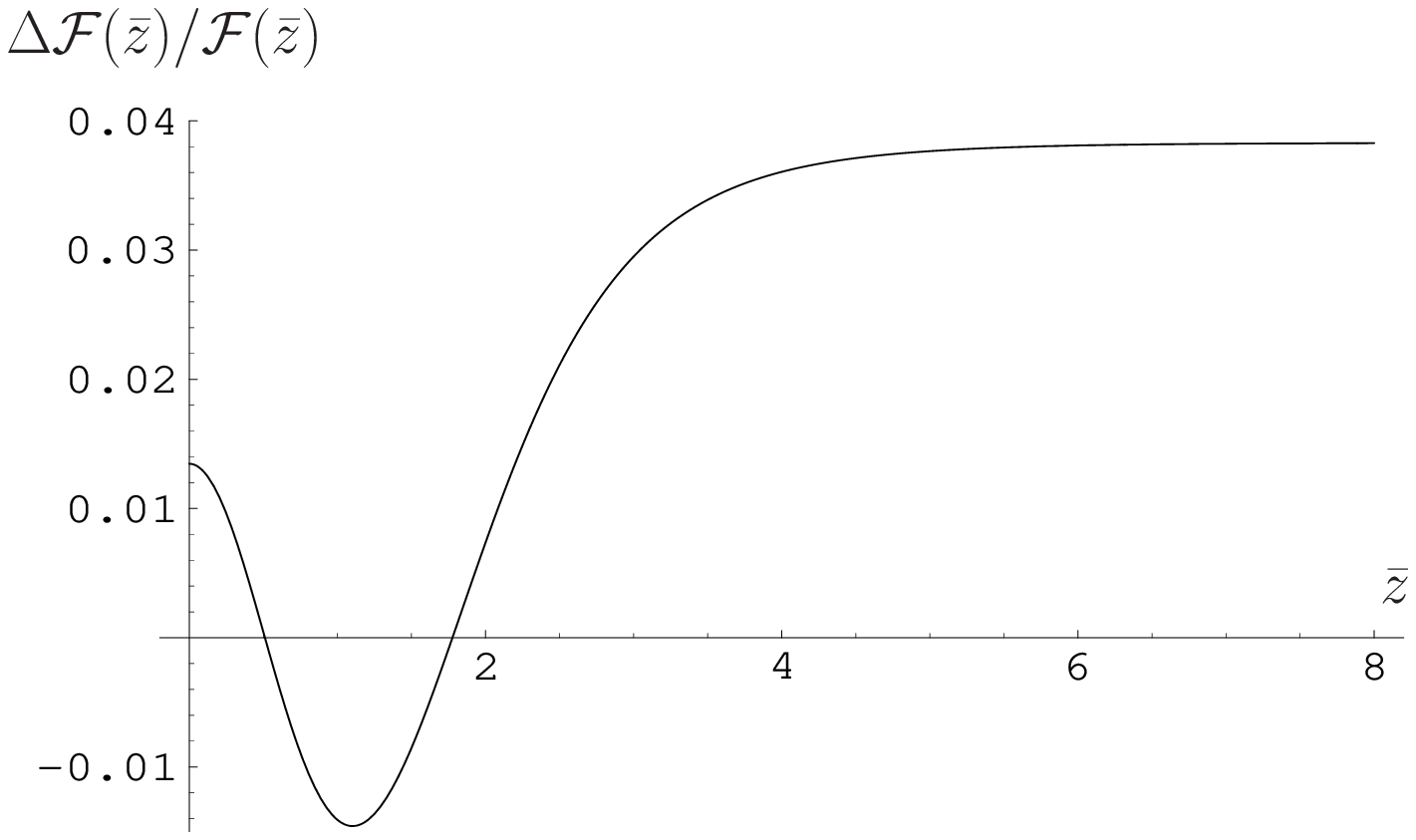}}
\caption[a]
{
Left: The domain wall free energy densities $\mathcal F(\zp)$,
divided by $\coeff 13\pi^2 T^4$,
in the effective theory and underlying Yang-Mills theory,
plotted versus $\bar z \equiv g(T) T z$.
The two curves are nearly indistinguishable,
differing by less than 1.5\% at the center of the domain wall.
Right: The relative difference of the two curves,
$(\mathcal F_{\rm eff}(\bar{z})-\mathcal F_{\rm YM}(\bar{z}))/\mathcal F_{\rm YM}(\bar{z})$.
\label {fig:wall}
}
}
\end{FIGURE}

A straightforward calculation shows that the effective theory
reproduces the domain wall tension and width of the underlying
Yang-Mills theory when
\begin{eqnarray}
    \vv&=&3.005868 \,,\qquad
    \tilde{c}_2 = 0.118914 \,. \label{ct2}
\end{eqnarray}
[Note that this value for $\bar v$, almost equal to 3,
makes the minima of the potential lie almost exactly at $T$
times an element of $Z(3)$.]
In Fig.~\ref{fig:soln} we plot $\alpha(\zp)$ for these values of the
parameters, and in Fig.~\ref{fig:wall}
we compare the resulting domain wall free energy density
in the effective theory, given by
\begin{equation}
    \mathcal F_{\rm eff}(z) = 
    (4\pi \vv \, T^2)^2 (\coeff 13)^3 \> \widetilde U(\alpha(\zp(z))) \,,
\label {eq:Feff}
\end{equation}
with the full theory result (\ref {eq:FYM}).
The agreement is rather remarkable.

\subsection {Discussion}

Perturbative matching with EQCD,
plus comparison of domain wall properties,
has determined (to leading order) all parameters of the
effective theory
[leading to Eqs.~(\ref{ct1}), (\ref{ct3}), and (\ref{ct2})]
except the two heavy masses $m_\phi$ and $m_\chi$.
These heavy masses are arbitrary, as long as they are
parametrically of order $T$ (or $v$), and satisfy $m_\phi > \coeff 13 m_\chi$.

It may seem surprising that domain wall properties can be
used to fix parameters of the effective theory
which were left undetermined by the perturbative matching to EQCD.
Correct matching with EQCD should imply that our
effective theory reproduces all physics of hot Yang-Mills theory on energy
scales small compared to $T$.
Fields in the domain wall only vary on the Debye screening length scale
$[g(T) T]^{-1}$ which, for asymptotically high temperatures,
is parametrically large compared to $T^{-1}$.
So why didn't matching to EQCD yield an effective theory
that automatically reproduces the correct domain wall structure,
without any further tuning of parameters?

The resolution of this puzzle involves the realization that the
inverse domain wall width is not the only relevant energy scale
in a domain wall.
As seen in the fluctuation contribution (\ref {eq:Ffluc}) to the tension,
near the center of the domain wall off-diagonal components of the static
gauge field receive effective masses which are order $v$ in the
effective theory, and order $T$ in the underlying Yang-Mills theory.%
\footnote
    {
    An alternative way to see this is to consider the
    electric field contribution to the Yang-Mills action,
    $
    \tr (D_0 A_i)^2
        =\tr (\partial_0 A_i -i g [A_0,A_i])^2
    $.
    A background value of $A_0$ effectively shifts the
    Matsubara frequencies from $2\pi n T$ to $2\pi n T$
    minus the eigenvalues of $g A_0$ (in the adjoint representation).
    Near the center of a domain wall,
    the eigenvalues of $A_0$ (in a gauge where $A_0$ is static)
    are of order $\pi T/g$ ---
    since this corresponds to an order one change in the phase
    of the Wilson line $\Omega \sim e^{ig \beta A_0}$.
    Consequently, near the center of a domain wall there is no
    separation of scales between the effective frequencies
    of the static and non-static modes.
    }
Hence, the $O(T)$ value of $v$, which determines the magnitude
of the scalar field in our effective theory, is a physically
relevant scale in the domain wall.
Equivalently, the excess free energy density (\ref {eq:Feff})
near the center of the domain wall is $O(T^4)$, and thus comparable
to the $O(T^4)$ equilibrium free energy density due to
fluctuations on the scale of $T$.

The symmetry structure of the effective theory
guarantees that it will have domain walls interpolating between
different $Z(3)$ minima, and the matching to EQCD guarantees
that these solutions will have the correct
exponential decay, proportional to $e^{-\mD |z|}$,
far from the domain wall.
However, the precise value of the domain wall tension,
and its width, are sensitive to physics on the scale of $T$.
This is why reproducing these observables in the effective theory
requires tuning of parameters beyond the perturbative matching
to EQCD.
If one were to demand that the effective theory exactly reproduce
all moments of the domain wall energy density (\ref {eq:FYM}),
then it would be necessary to add an infinite number of additional
higher dimension operators to the effective theory, with
appropriately tuned coefficients.
But from Fig.~\ref{fig:wall}, it is evident that near perfect agreement
is achieved, without introducing any higher dimension operators,
just by suitably adjusting those low energy parameters ($v$ and $\tilde c_2$)
of our superrenormalizable effective theory which were left undetermined
by the matching to EQCD.

To conclude this section, let us briefly review the different sources of
corrections to the Lagrangian and parameters of the effective theory. These will become
important if one aims at next-to-leading-order (NLO)
accuracy in the determination of various physical quantities
through the effective theory, such as bulk equilibrium thermodynamics or the $Z(3)$ domain
wall tension and width.

Integrating out the heavy fields,
with vertices originating from the hard potential $V_0(\Z)$,
will produce corrections to the scalar and gauge field kinetic terms.
At first non-trivial order, the former involves determining the leading momentum dependence
of the two-point graphs of Fig.~\ref{fig:match}b,
and will yield an order $g^2$ correction to the normalization of the scalar
kinetic term in the Lagrangian of Eq.~(\ref{lageff3}).%
\footnote
    {
    Terms involving powers of the external momentum,
    which correspond to higher derivative terms in the Lagrangian,
    will be proportional to inverse powers of the heavy masses and
    are thus suppressed by additional factors of $g$ when evaluated on
    momentum scales of order $gT$ which are relevant in the effective theory.
    }
These corrections amount to multiplicative wave function renormalizations
for the fields in question.

Integrating out heavy fields at NLO will also generate
corrections to the values of the parameters in the light theory
(\ref{lageff3}),
whose matching to EQCD will require $\mathcal O(g^2)$ relative
adjustments to the couplings in the soft potential $V_1(\Z)$.
For the parameters $m_a$ and $\tilde \lambda$, one-loop
contributions involving one soft vertex will
produce ${\mathcal O}(g^4)$ and ${\mathcal O}(g^6)$
corrections, respectively.
For bulk thermodynamics, the NLO correction to $m_a^2$ is more
important than the leading order value of $\tilde \lambda$.

Because the $SU(3)\times SU(3)$ invariance of $V_0$
is not respected by the soft potential $V_1$, or by the coupling of
the scalar fields to the gauge field, integrating out the heavy fields
may also produce entirely new operators involving $\mathbf A$ and $a$
of dimension three or higher.
The first such new operators to appear
include $\tr a^6$ and $\tr (a^2(D_i a)^2)$,
and are suppressed by $\gE^8$.%
\footnote
    {
    For  $\tr a^6$, one factor of $\gE^2$ comes from the necessary
    inclusion of one soft vertex from $V_1$.
    For $\tr (a^2(D_i a)^2)$, the multiplicative coupling constant is of order
    $g_3^6$, but the covariant derivatives contribute additional factors
    of $g_3$ --- either explicitly or via spatial derivatives
    applied to $gT$ scale physics.
    The same is true for other higher dimension operators
    involving derivatives.
    }

Ensuring that the domain wall properties of our theory agree with
corresponding results in full Yang-Mills theory beyond leading order
would, in addition to the above, require extending the derivative expansion of the one-loop fluctuation
contribution $F_{\rm fluc}$ to next-to-leading order, as well
as evaluating the two-loop effective potential for $\Z$.
The analogous NLO calculation in the full Yang-Mills theory is discussed
in more detail in Ref.~\cite{kap2}. The required analysis in the
effective theory should be very similar.

\section{Phase diagrams}

\subsection {EQCD}

The lack of $Z(3)$ center symmetry in EQCD
has been a primary motivation for the work presented in this paper.
The consequences of this explicit breaking of the $Z(3)$ symmetry
are particularly apparent if one studies the phase diagram of EQCD
for arbitrary values of its parameters, as done in Refs.~\cite{klrrt,klrsh}.
One can form two dimensionless ratios
from the parameters ($\gE^2$, $\mE^2$, and $\lambda_{\rm E}$) of EQCD,
conventionally chosen to be%
\footnote
    {
    Our definition of $\lambda_{\rm E}$ differs from that of Ref.~\cite{klrrt}
    by a factor of $2/\gE^2$.
    }
$x\equiv\half\lambda_\rmi{E}$ and $y\equiv m_\rmi{E}^2/\gE^4$.
Parameter values which result from matching to hot $SU(3)$ Yang-Mills theory
correspond to the curve
\begin{equation}
    xy = \frac 3{8\pi^2} + \frac 9{16 \pi^2} \, x + \mathcal O(x^2) \,,
\label {eq:DR}
\end{equation}
where the next-to-leading order correction, proportional to $x$,
has been included.
Increasing temperature corresponds to decreasing values of $x$
(proportional to $g(T)^2$).

\begin{FIGURE}[ht]
{
\centerline{\def\epsfsize#1#2{0.65#1}\epsfbox{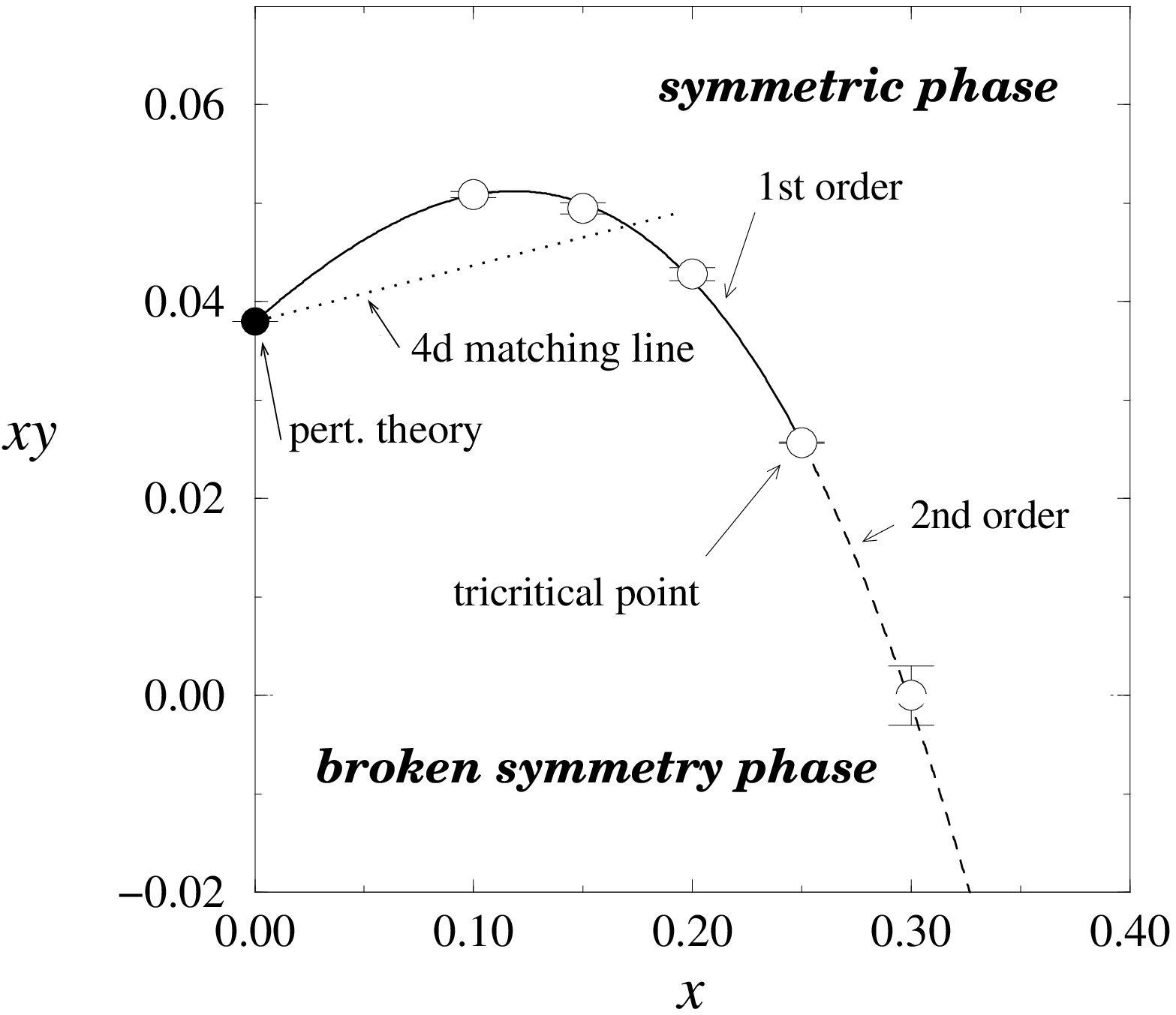}}
\vspace*{-2em}
\caption[a]{The phase diagram of EQCD (from Ref.~\cite{klrrt})
parametrized by
$x\equiv\half\lambda_\rmi{E}$ and $y\equiv m_\rmi{E}^2/\gE^4$.
The phase with spontaneous symmetry breaking of the $A_0 \to -A_0$ symmetry
lies in the lower left region of the diagram;
the upper region is the unbroken symmetry phase.
The dotted ``$4d$ matching line'' is the slice of the phase diagram
which satisfies relation (\ref {eq:DR}) giving the result of
next-to-leading order matching with hot $SU(3)$ Yang-Mills theory.
\label{fig:phase}
}
}
\end{FIGURE}

For sufficiently small values of $\mE^2$ (or $y$)
one finds that the $A_0 \to -A_0$ symmetry,
which is a remnant of time reflection symmetry
in the original 4-$d$ theory, is spontaneously broken.
Within the broken phase, there are two degenerate equilibrium states,
with opposite (and non-zero) expectation values for $\tr A_0^3$.
The phase transition line separating the symmetric phase
(with vanishing $\langle \tr A_0^3 \rangle$)
from this symmetry broken phase is first order for small values of $x$
and second order for large values, with a tricritical point
at $x \approx 0.26$.
This is illustrated in Fig.~\ref{fig:phase}.
On the first order part of the transition line there are three
co-existing extremal equilibrium states,
distinguished by positive, negative, and zero values of
the order parameter $\langle \tr A_0^3 \rangle$.

Near the first order transition line,
these three equilibrium states with differing values of
$\langle \tr A_0^3\rangle$
may be regarded as deformed remnants of the three
equilibrium states
of hot $SU(3)$ gauge theory
(which are related by the center symmetry).
But in EQCD, there is no three-fold symmetry relating these states,
only a $Z(2)$ symmetry interchanging the two symmetry-broken states.
The states with non-zero $\langle \tr A_0^3\rangle$ have,
for example, energy densities and correlation lengths which differ
from the symmetric $\langle \tr A_0^3\rangle = 0$ state \cite{klrrt}.

Only the symmetric phase corresponds to the physics of the underlying
hot Yang-Mills theory.
However, as indicated in Fig.~\ref{fig:phase},
the dimensional reduction line (\ref {eq:DR}) lies inside the broken phase!
This line intersects the phase transition curve at $x = 0$
(or $T = \infty$),
but for any non-zero value of $x$ (or finite temperature $T$),
matching with hot Yang-Mills theory requires studying the
symmetric phase on the wrong side of the phase transition ---
that is, studying a metastable ``supercooled'' phase.
For sufficiently small $x$, this is feasible.
The lifetime of the metastable phase is so long that transitions
to the stable symmetry broken phase are never seen in practical
numerical simulations.
But this is surely a less-than-satisfying feature of EQCD.

As one increases $x$ (or lowers the temperature) in EQCD,
there is no region where a new ``confining'' phase appears,
and no quadruple point where four equilibrium states coexist,
analogous to the transition temperature $T_c$ in real $SU(3)$
Yang-Mills theory, where the three spontaneously broken $Z(3)$
deconfined states and the $Z(3)$ symmetric confining state all coexist.

\subsection {$Z(3)$ invariant effective theory}

Viewed as a superrenormalizable three dimensional field theory
with no constraints on its coupling constants (besides stability),
our $Z(3)$ invariant effective theory (\ref {lageff2}--\ref{v1})
depends on a total of seven parameters,
which may be taken to be one overall scale $v$,
the ratios of the heavy masses to this scale
($m_\phi^2/v^2$ and $m_\chi^2/v^2$),
a dimensionless gauge coupling $\gE^2/v$,
and three dimensionless ``soft'' couplings
($\tilde c_1/v$, $\tilde c_2$, and $\tilde c_3 \, v$).
Exploring the resulting phase diagram via numerical simulations,
as all six dimensionless ratios are varied, is obviously impractical.
However, it is inevitable that the parameter space of this theory contains
(i) a region in which the $Z(3)$ symmetry is spontaneously broken, and
(ii) a region in which the $Z(3)$ symmetry is unbroken.%
\footnote
    {
    Gauge symmetries are, of course, never truly spontaneously broken.
    In addition to the global $Z(3)$ symmetry,
    our effective theory is also invariant under $\Z \to \Z^*$
    (for real values of the couplings).
    Hence, the global symmetry group is really $Z(3) \times Z(2) = Z(6)$.
    The $Z(3)$ broken phase is a phase in which this $Z(6)$
    global symmetry spontaneously breaks to a $Z(2)$ subgroup.
    For real values of the couplings,
    we do not believe that any other symmetry breaking pattern
    (such as $Z(6) \to {}$nothing) is possible.
    }

There are two ways in which one can easily imagine changing the parameters
of the theory so as to move from the symmetry broken
(deconfined, high temperature)
phase to the unbroken (confining, low temperature) phase.
If the heavy masses are varied and $m_\phi$ becomes less than
$\coeff 13 m_\chi$, then the global minimum of the potential $V_0(\Z)$
switches from $\Z = v\openone$ (times a cube root of unity) to the symmetric
point $\Z = 0$.%
\footnote
    {
    Within a mean field approximation, $m_\phi$ and $m_\chi$ are the
    inverse correlation lengths for the real and imaginary parts
    of the trace of the Wilson line, respectively.
    A ratio of three for $m_\chi/m_\phi$ at the point where the
    tree-level potential undergoes symmetry restoration is in
    accord with the mean field analysis of Dumitru and Pisarski \cite{dumpis}
    for a $Z(3)$ symmetric model involving only the trace of the Wilson line
    (with up to quartic interactions).
    }
If the theory is kept weakly coupled, so that $\gE^2/v \ll 1$,
then this tree-level (or mean field) analysis is reliable,
and the resulting transition will generically be a strong first order
transition.

Alternatively, if $m_\phi$ remains greater than $\coeff 13 m_\chi$
but the theory becomes sufficiently strongly coupled,
so that $\gE^2/v$ is no longer small,
then this should drive a fluctuation induced phase transition
in which the entropy-driven effects of fluctuations overwhelm
the symmetry-breaking influence of the potential.
This is analogous to symmetry restoration in non-linear sigma models.

    The confining/deconfining phase transition in $SU(3)$ Yang-Mills
theory is known to be a relatively weak first order transition,
and near the transition fluctuations of the Wilson line,
or the static gauge field, are not weakly coupled.
Consequently, we expect that the second regime of fluctuation induced
phase transitions is the more appropriate model for $SU(3)$ gauge theory
near its confinement transition.

An interesting two-dimensional slice of the parameter space
of our effective theory which would probe both of the above regimes,
as well as the portion of parameter space which corresponds to
asymptotically high temperature Yang-Mills theory,
is given by varying $\gE^2/v$ and $m_\phi^2/v^2$
while fixing the soft couplings at the values determined
from the high temperature matching in the previous section, namely
\begin {equation}
    \tilde c_1/v = 0.332683 \,,\qquad
    \tilde c_2 = 0.118914 \,,\qquad
    \tilde c_3 \, v = 0.228419 \,,
\end {equation}
and choosing $m_\chi^2/v^2$ to equal, say, unity.
Within this two-dimensional slice, there will be a region of
unbroken $Z(3)$ symmetry,
a region of spontaneously broken $Z(3)$ symmetry,
and a phase transition line separating these regions.
The entire phase transition line is expected to be first order
(for the same reasons discussed in Ref.~\cite {ys}),
and everywhere on the transition line there will be four co-existing
equilibrium states --- just as in the full $SU(3)$ gauge theory.

Exploring this two-dimensional parameter space
via numerical simulations is surely feasible.
As a warm-up, one could focus on a single line with a
fixed value $m_\phi^2/v^2$, and only vary $\gE^2/v$.
It would be very interesting to see if the behavior of physical observables
(such as the ratios of correlation lengths in different symmetry channels)
along such a line closely mimics the behavior in hot $SU(3)$ gauge theory
as the temperature varies from below $T_c$ to far above.

\section{Conclusion}

We have formulated a three-dimensional $Z(3)$-invariant effective theory
which can reproduce properties of hot $SU(3)$ Yang-Mills theory both
at asymptotically high temperatures,
and near the confinement/deconfinement transition.
Lowest order perturbative matching plus comparison of domain wall
properties was used to fix the parameters of the effective theory
at high temperatures.
The form of the effective theory ensures that
as one increases the dimensionless gauge coupling $\gE^2/v$
(which increases the size of fluctuations),
or decreases the heavy mass $m_\phi$
(which changes the shape of the tree-level potential),
the theory will undergo a first order phase transition from a
deconfined phase with spontaneously broken $Z(3)$ symmetry
to a confining phase with unbroken $Z(3)$ symmetry.
This reproduces, by design, the qualitative behavior of
hot $SU(3)$ Yang-Mills theory near $T_c$.
Numerical simulations of the effective theory will be needed
to determine the tuning of parameters that makes the effective theory,
near its phase transition, most closely reproduce quantitative properties
of hot Yang-Mills theory near the confinement transition.
As discussed in the previous section, just varying $\gE^2/v$ while
holding the other parameters fixed at the values determined by
high temperature matching will give a theory that reproduces
the correct high temperature behavior at small gauge coupling,
together with a fluctuation induced phase transition at a critical
value of $\gE^2/v$.
This, we hope, will be an accurate effective theory for quarkless QCD
for temperatures ranging from near $T_c$ to asymptotically large.
We look forward to numerical simulations testing this hypothesis.%
\footnote
    {
    Such simulations are currently underway \cite{Aleksi2}.
    }

In the case of a positive result,
it will be interesting to investigate generalizations of our effective
theory to other values of $N_\rmi{c}$, or to QCD with dynamical quarks.
For sufficiently heavy quarks, it is known that a weakly first order
finite temperature phase transition persists, even though fundamental
representation quarks break the $Z(3)$ center symmetry and no local
order parameter for the transition is available.
Therefore, dynamical quarks may be viewed as adding soft $Z(3)$
symmetry breaking terms to the effective theory.
Such terms will lift the degeneracy of the minima of the potential,
leaving the minimum with a real expectation value for the trace of
the Wilson line as the unique global minimum.
Matching, at high temperature, to EQCD with a non-zero number of quark flavors
should be straightforward.
The resulting effective theory should capture
some of the $Z(3)$ physics postulated to be important
in the full theory even at non-zero $N_\rmi{f}$ \cite{pisa}.
Pursuing this generalization will be left for future work.

\section*{Acknowledgments}
We gratefully acknowledge useful discussions with Keijo Kajantie
and Mikko Laine, especially regarding the phase diagram of EQCD.
This work was supported by the U.S.~Department of Energy
under Grant No.~DE-FG02-96ER40956.

\newpage

\appendix

\section{Minimizing the potential}\label{sec:potmin}

The variation of the hard potential $V_0(\Z)$, defined in Eq.~(\ref{v0}), is
\begin{eqnarray}
    \delta V_0
    &=&
    \tr\!\left[
    \delta \Z
    \left\{
        c_1 \, \Z^{\dagger}
        +c_2 \, \Z^{-1}\det \Z+
        2c_3 \, \Z^{\dagger}\Z \, \Z^{\dagger}
    \right\}
    \right]
    +
    \mbox{h.c.}
\end{eqnarray}
and this vanishes when $\Z$ satisfies
\begin{equation}
    c_1 \, \Z^{\dagger}\Z
    +c_2 \, \det \Z
    +2c_3 \, (\Z^{\dagger}\Z)^2
    = 0 \,.
\label{detcond}
\end{equation}
Subtracting this equation from its Hermitian conjugate shows that
$\det\Z$ must be real at extrema.
For $c_2 < 0$, which we have assumed,
one may see directly from the potential (\ref{v0})
that the minimum of $V_0(\Z)$ will occur
for matrices with positive determinant.

Let $\lambda^2$ denote a (positive, real) eigenvalue of the
the positive-definite matrix $\Z^{\dagger}\Z$.
To satisfy relation (\ref{detcond}), these eigenvalues
must obey
\begin{eqnarray}
    2c_3 \, \lambda^4 + c_1 \, \lambda^2 + c_2 \, \det \Z&=&0 \,.
\end{eqnarray}
For $c_2 < 0 < c_3$, and a positive determinant of $\Z$,
this has a unique positive root,
\begin{eqnarray}
    \lambda^2 &=& \fr{-c_1 + \sqrt{c_1^2-8c_2c_3\det \Z}}{4c_3}.
\label{eq:lambda}
\end{eqnarray}
Therefore, all eigenvalues of $\Z^\dagger \Z$ are equal,
implying that this matrix is proportional to the unit matrix.
Consequently, $\Z$ must equal a special unitary matrix times $\lambda$.
So $\det\Z = \lambda^{3}$ and condition (\ref {eq:lambda}) becomes
$
    \lambda^2 = [-c_1+\sqrt{c_1^2-8c_2c_3\lambda ^3}]/(4c_3)
$.
The solutions of this equation are $\lambda=0$ and
$\lambda = \lambda_\pm$ where
\begin{eqnarray}
    \lambda_\pm &\equiv&  \fr{-c_2\pm \sqrt{c_2^2-8c_1c_3}}{4c_3} \,.
\label{vroot}
\end{eqnarray}
These non-zero roots are real if $c_2^2 > 8 c_1 c_3$.
Evaluating $V_0$ at its three possible extrema,
one finds
that the root $\lambda_+$ corresponds to a global minimum provided
$c_2 < 0$ and $c_2^2 > 9 c_1 c_3$.
Assuming $c_2$ satisfies these conditions,
$V_0(\Z)$ is minimized
when $\Z = \lambda_+ \, \Omega$ for arbitrary $\Omega \in SU(3)$.
When $c_1 > 0$ there is a local minimum at $\Z = 0$ and
the condition $c_2^2 > 9 c_1 c_3$ implies that
$c_2$ must be less than $-3 \sqrt{c_1 c_3}$.
But if $c_1 < 0$ then $\Z = 0$ is a local maximum and
$c_2$ must merely be negative.
If $c_1 > 0$ and $8 c_1 c_3 < c_2^2 < 9 c_1 c_3$,
then $\Z = 0$ is the global minimum of the potential and
$\Z = \lambda_+ \, \Omega$
is only a local minimum.

To analyze the soft potential $V_1(\Z)$,
given in Eq.~(\ref{v1}), it is convenient
to insert the singular value decomposition
$M = L m R^\dagger$.
Here $L$ and $R$ are unitary matrices, while $m$ is
diagonal, real and positive.
Only the cubic terms in $V_1(\Z)$ depend on the unitary matrices $L$ and $R$,
and it is easy to see that
$\tilde c_2 \left( \tr[M^3] + \tr[(M^\dagger)^3] \right)$
is minimized for $L = R$ if $\tilde c_2 < 0$,
or $L = -R$ if $\tilde c_2 > 0$.
Therefore, it is sufficient to minimize $V_1(\Z)$ for
$M$ a diagonal real matrix.
This gives
\begin{equation}
    V_1(\Z) = \sum_i \>
    \(\tilde c_1 \, \mu_i^2
    + 2\tilde c_2 \, \mu_i^3
    + \tilde c_3 \, \mu_i^4\) ,
\end{equation}
where $\{\mu_i \}$ are the real eigenvalues of $M$.
Solving for zeros of this expression, one finds that $V_1(\Z)$
vanishes only at $M = 0$, and is strictly positive for all non-zero $M$,
provided
$\tilde c_1$ and $\tilde c_3$ are positive and
$\tilde{c}_1\tilde{c}_3>\tilde{c}_2^{\,2}$, which we assume.

Non-trivial minima of the hard potential $V_0(\Z)$
[{\em i.e.}, $\Z / \lambda_+ \in SU(3)$]
coincide with minima of the soft potential $V_1(\Z)$
[{\em i.e.}, configurations satisfying $\Z \propto \openone$ so that $M = 0$]
when $\Z$ equals $\lambda_+$ times an element of the center of $SU(3)$.
So the global minima of the complete potential
$V_0(\Z) + \gE^2 \, V_1(\Z)$
(when $c_2^2 > 9 c_1 c_3$)
lie at
\begin{eqnarray}
    \Z&=& e^{2\pi in/3} \, \lambda_+ \,\openone\,,\qquad n\in\{0,1,2\} \,.
\end{eqnarray}

Using the decomposition (\ref{eq:decomp}) of $\Z$ and the resulting
relations (\ref {c1}) between the coefficients
in $V_0(\Z)$ and the tree level heavy masses,
one may translate the above conditions on the coefficients
into equivalent conditions on $m_\phi$ and $m_\chi$
(which are the tree level masses for fluctuations in the
magnitude and phase of $\tr \Z$, respectively).
The results may be summarized as follows.

\begin{enumerate}
\item
    If $m_\chi^2 > 0$ and $m_\chi^2 + 3m_\phi^2 > 0$ then
    $c_2 < 0 < c_3$.
    These conditions are required so that $V_0(\Z)$ is bounded below,
    and favors matrices with positive determinant.
\item
    If $m_\phi^2 > m_\chi^2/3$ then $c_1$ is negative
    and there is a local maximum at $\Z = 0$ and three degenerate
    global minima at $\Z = \lambda_+ \, e^{2\pi in/3} \openone$.
\item
    If $m_\chi^2 / 9 < m_\phi^2 < m_\chi^2/3$ then
    $c_1 > 0$ and $c_2 < -3 \sqrt{c_1 c_3}$, implying that
    there is a local minimum at $\Z = 0$ and three degenerate
    global minima at $\Z = \lambda_+ \, e^{2\pi in/3} \openone$.
\item
    If $0 < m_\phi^2 < m_\chi^2/9$ then
    $c_1 > 0$ and $-3 \sqrt{c_1 c_3} < c_2 < -\sqrt{8 c_1 c_3}$, in which case
    there is a global minimum at $\Z = 0$ and three degenerate
    local minima at $\Z = \lambda_+ \, e^{2\pi in/3} \openone$.
\end{enumerate}

\section{Expansion in shifted fields}\label{sec:potshift}

Expressing the potential (\ref {eq:V}) in terms of the
shifted fields (\ref{eq:decomp}) which describe fluctuations
away from the minimum gives
\begin {eqnarray}
    \gE^{-2} \, V(\Z)
    &=&
    V_{\rm min}
    + \half m_\phi^2 \> \phi^2
    + \half m_\chi^2 \> \chi^2
    + m_h^2 \> \tr\!( h^2 )
    + V_{\rm int}(\phi,\chi,h,a) \,,
\\
\noalign{\hbox{with}}
    V_{\rm min} &=& \coeff 1{54} \, (9c_1+vc_2) \, v^2 / \gE^2 \,,
\end{eqnarray}
and
\begin{align}
    V_{\rm int}(\phi,\chi,h,a)
    &=  
    -\fr {g_3}{\sqrt 6 \, v}
    \Bigl\{
        (3c_1+\coeff 23 vc_2) \, \phi^3
        + (3c_1+2vc_2) \, \phi\chi^2
        + (6c_1+vc_2) \, \phi \, \tr (a^2)
\nn & \quad\qquad\qquad{}
        + (18c_1+7vc_2) \, \phi \, \tr (h^2)
        + (12c_1+2vc_2) \, \chi \, \tr (ah)
\nn & \quad\qquad\qquad{}
        + \sqrt 6 \, (6c_1+4vc_2) \, \tr(a^2h)
        - \sqrt 6 \, (6 c_1+ \coeff 43 vc_2) \, \tr (h^3)
    \Bigr\}
\nn & {}
    - \fr{g_3^2}{v^2} \, ( 3c_1+vc_2 )
    \Bigl\{
        \coeff{1}{8} \(\phi^2+\chi^2\)^2
        + \coeff{1}{2} \phi^2 \, \tr ( a^2+3 h^2 )
        + 2 \phi\chi \, \tr(ah)
\nn & \quad\qquad\qquad\qquad{}
        + \coeff{1}{2} \chi^2 \, \tr ( 3 a^2+ h^2 )
        +2 \phi \, \tr ( a^2h+ h^3 )
        +2 \chi \, \tr ( a^3+ ah^2 )
\nn & \quad\qquad\qquad\qquad{}
        + \coeff{3}{2} \, \tr ( a^4 + 4 a^2h^2 -2(ah)^2+ h^4 )
    \Bigr\}
\nn & {}
    + g_3^2 \big\{ \tilde{c}_1 \, \tr ( a^2+ h^2 ) \big\}
    - g_3^3 \big\{ 2\tilde{c}_2 \, \tr(3a^2h- h^3 ) \big\}
\nn[5pt] & {}
    + g_3^4 \big\{ \tilde{c}_3 \, \tr (a^4+4 \, a^2h^2-2(ah)^2+ h^4) \big\} ,
\label{eq:Vshift}
\end{align}
Here $c_3$ has been expressed in terms of $v$ using Eq.~(\ref{eq:v}).
The tree level masses are
\begin{equation}
    m_{\phi}^2 = -2 c_1 - \coeff 13 v \, c_2 \,,\qquad
    m_{\chi}^2 = -v  \, c_2 \,,\qquad
    m_{h}^2 = -2 c_1 - \, \coeff 43 v \, c_2 \,.
\end{equation}
Expressed in terms of these masses,
$
    V_{\rm min} = -\fr{v^2}{108}(9m_{\phi}^2-m_{\chi}^2)/\gE^2
$.
This shows directly that $m_\phi^2 > \coeff 19 m_\chi^2$
is necessary for the global minima to lie at
the non-trivial $Z(3)$ extrema.



\begin{thebibliography}{99}

\bibitem{bn1}
E.~Braaten and A.~Nieto,
``Free energy of QCD at high temperature,''
Phys.\ Rev.\ D {\bf 53} (1996) 3421
[hep-ph/9510408].

\bibitem{klry}
K.~Kajantie, M.~Laine, K.~Rummukainen and Y.~Schr\"oder,
``The pressure of hot QCD up to $g^6 \ln(1/g)$,''
Phys.\ Rev.\ D {\bf 67} (2003) 105008 [hep-ph/0211321].

\bibitem{hlp2}
  A.~Hart, M.~Laine and O.~Philipsen,
  ``Static correlation lengths in QCD at high temperatures and finite
  densities,''
  Nucl.\ Phys.\ B {\bf 586} (2000) 443
  [hep-ph/0004060].

\bibitem{lv}
  M.~Laine and M.~Veps\"al\"ainen,
  ``Mesonic correlation lengths in high-temperature QCD,''
  JHEP {\bf 0402} (2004) 004
  [hep-ph/0311268].

\bibitem{klry3}
  K.~Kajantie, M.~Laine, K.~Rummukainen and Y.~Schr\"oder,
  ``How to resum long-distance contributions to the QCD pressure?,''
  Phys.\ Rev.\ Lett.\  {\bf 86} (2001) 10
  [hep-ph/0007109].

\bibitem{klrrt}
K.~Kajantie, M.~Laine, A.~Rajantie, K.~Rummukainen and M.~Tsypin,
``The phase diagram of three-dimensional $SU(3)$ + adjoint Higgs theory,''
JHEP {\bf 9811} (1998) 011
[hep-lat/9811004].


\bibitem{klry2}
K.~Kajantie, M.~Laine, K.~Rummukainen and Y.~Schr\"oder,
``Four-loop logarithms in 3D gauge + Higgs theory,''
Nucl.\ Phys.\ Proc.\ Suppl.\  {\bf 119} (2003) 577
[hep-lat/0209072].


\bibitem{pisa}
R.~D.~Pisarski,
``Quark-gluon plasma as a condensate of $SU(3)$ Wilson lines,''
Phys.\ Rev.\ D {\bf 62} (2000) 111501
[hep-ph/0006205].


\bibitem{kovner}
A.~Kovner,
``Confinement, magnetic $Z(N)$ symmetry and low-energy effective theory of gluodynamics,''
hep-ph/0009138.



\bibitem{peter}
P.~Bialas, A.~Morel and B.~Petersson,
``A gauge theory of Wilson lines as a dimensionally reduced model of  $QCD_3$,''
Nucl.\ Phys.\ B {\bf 704} (2005) 208
[hep-lat/0403027].

\bibitem{wiese}
K.~Holland and U.~J.~Wiese,
``The center symmetry and its spontaneous breakdown at high temperatures,''
hep-ph/0011193.

\bibitem{ay}
  P.~Arnold and L.~G.~Yaffe,
  ``The non-Abelian Debye screening length beyond leading order,''
  Phys.\ Rev.\ D {\bf 52} (1995) 7208
  [hep-ph/9508280].

\bibitem{owe}
  O.~Philipsen,
  ``On the non-perturbative gluon mass and heavy quark physics,''
  Nucl.\ Phys.\ B {\bf 628} (2002) 167
  [hep-lat/0112047].


\bibitem{lp}
  M.~Laine and O.~Philipsen,
  ``The non-perturbative QCD Debye mass from a Wilson line operator,''
  Phys.\ Lett.\ B {\bf 459} (1999) 259
  [hep-lat/9905004].

\bibitem{kacz}
  O.~Kaczmarek, F.~Karsch, E.~Laermann and M.~Lutgemeier,
  ``Heavy quark potentials in quenched QCD at high temperature,''
  Phys.\ Rev.\ D {\bf 62} (2000) 034021
  [hep-lat/9908010].


\bibitem{su3order}
  M.~Fukugita, M.~Okawa and A.~Ukawa,
  ``Finite size scaling study of the deconfining phase transition in pure
  $SU(3)$ lattice gauge theory,''
  Nucl.\ Phys.\ B {\bf 337}, 181 (1990).


\bibitem{gpy}
D.~J.~Gross, R.~D.~Pisarski and L.~G.~Yaffe,
``QCD and instantons at finite temperature,''
Rev.\ Mod.\ Phys.\  {\bf 53} (1981) 43.

\bibitem{kap}
T.~Bhattacharya, A.~Gocksch, C.~Korthals Altes and R.~D.~Pisarski,
``Interface tension in an $SU(N)$ gauge theory at high temperature,''
Phys.\ Rev.\ Lett.\  {\bf 66} (1991) 998.


\bibitem{ys}
L.~G.~Yaffe and B.~Svetitsky,
``First order phase transition in the $SU(3)$ gauge theory at finite
temperature,''
Phys.\ Rev.\ D {\bf 26} (1982) 963.


\bibitem{bn0}
  E.~Braaten and A.~Nieto,
  ``Effective field theory approach to high temperature thermodynamics,''
  Phys.\ Rev.\ D {\bf 51}, 6990 (1995)
  [hep-ph/9501375].


\bibitem{kap2}
T.~Bhattacharya, A.~Gocksch, C.~Korthals Altes and R.~D.~Pisarski,
``$Z(N)$ interface tension in a hot $SU(N)$ gauge theory,''
Nucl.\ Phys.\ B {\bf 383} (1992) 497
[hep-ph/9205231].

\bibitem{klrsh}
K.~Kajantie, M.~Laine, K.~Rummukainen and M.~E.~Shaposhnikov,
``3d $SU(N)$ + adjoint Higgs theory and finite-temperature QCD,''
Nucl.\ Phys.\ B {\bf 503} (1997) 357
[hep-ph/9704416].


\bibitem{dumpis}
A.~Dumitru and R.~D.~Pisarski,
``Two-point functions for $SU(3)$ Polyakov loops near $T_c$,''
Phys.\ Rev.\ D {\bf 66} (2002) 096003
[hep-ph/0204223].


\bibitem{Aleksi2}
K.~Kajantie and A.~Kurkela, private communication.


\end{thebibliography}
\end{document}